\documentclass[journal]{IEEEtran}

\usepackage{xcolor,soul,framed} 

\colorlet{shadecolor}{yellow}
\usepackage[pdftex]{graphicx}
\graphicspath{{../pdf/}{../jpeg/}}
\DeclareGraphicsExtensions{.pdf,.jpeg,.png}

\usepackage[cmex10]{amsmath}
\usepackage{array}
\usepackage{mdwmath}
\usepackage{mdwtab}
\usepackage{eqparbox}
\usepackage{url}
\usepackage{algorithm}
\usepackage{algorithmicx}
\usepackage{algcompatible}
\usepackage{mathrsfs}
\usepackage{amssymb} 
\usepackage{booktabs}
\usepackage{mathtools}
\usepackage{mathrsfs}
\usepackage{multirow}
\usepackage{tabularx} 
\usepackage{dutchcal}
\usepackage{enumerate}
\usepackage[hidelinks]{hyperref}
\hypersetup{
	pdfauthor={Farid Zaredar, Morteza Amini},     
	pdftitle={A LightWeight Incentive-Based Privacy-Preserving Smart Metering Protocol for Value-Added Services}, 
	pdfsubject={Smart Metering Privacy},   
	pdfkeywords={Smart Grids, Smart Metering Privacy, Local Differential Privacy, Blind Digital Signatures, Hash-Chain Functions, Anonymous Overlay Network, Incentive-Based Value-Added Services}, 
	pdfcreator={LaTeX with hyperref}, 
	pdfproducer={pdflatex},     
	bookmarks=true,            
	bookmarksopen=true,        
	bookmarksnumbered=true,    
	colorlinks=false,          
}

\begin{document}
\bstctlcite{IEEEexample:BSTcontrol}
    \title{A Lightweight Incentive-Based Privacy-Preserving Smart Metering Protocol for Value-Added Services}
  \author{
  		Farid~Zaredar, and 
      	Morteza~Amini

		\thanks{F. Zaredar is with the Data and Network Security Laboratory (DNSL), Department of Computer Engineering, Sharif University of Technology, Tehran, Iran (e-mail: farid.zaredar78@sharif.edu).}
	  	
		\thanks{M. Amini is with Department of Computer Engineering, Sharif University of Technology, Tehran, Iran (e-mail: amini@sharif.edu).}

}

\markboth{}{Roberg \MakeLowercase{\textit{et al.}}: High-Efficiency Diode and Transistor Rectifiers}

\maketitle

\begin{abstract}
The emergence of smart grid and advanced metering infrastructure (AMI) has revolutionized energy management. AMI has enabled a wide variety of demand- and supply-side management utilities. Services such as accurate billing, outage detection, real-time grid control and monitoring, load forecasting, and value-added services can be considered as essential utilities in the energy infrastructure. In AMI, meters deliver consumption values at predefined intervals to the utility provider (UP). Reports of such readings can raise privacy violation issues. With such data, an adversary can deduce information about individuals' life patterns, political orientations, and the types of electrical and smart devices within their homes, or even sell the data to third parties (TP) such as insurance companies. In this paper, we propose a lightweight, privacy-preserving smart metering protocol to provide incentive-based value-added services. The scheme employs temporal-based aggregation and local differential privacy to enable reporting of coarse-grained consumption values with adjustable granularity while limiting the information revealed to the UP. Hash-chain credentials and blind digital signatures enable anonymous authentication of reported readings, while pseudonyms and anonymous overlay networks conceal the source of reports, providing anonymity and unlinkability between customers and their reported consumption values.
 This approach not only protects consumers' privacy but also preserves data utility. We demonstrate that the scheme prevents identity disclosure while enabling automatic token redemption at the specified time. Finally, we evaluated our protocol from two aspects: performance and privacy. From a performance perspective, our results show that with a 2048-bit RSA key size, a 7-day program duration, and four reports per day, our complete protocol implementation across the experimental setup takes about 3.36 seconds and consumes an average of 4.5 MB of memory during execution. From a privacy perspective, our formal analysis demonstrates consumption-value privacy, participant anonymity, and issuance–reporting unlinkability against semi-trusted aggregators and utility providers, while resisting untrusted active and passive adversaries.
\end{abstract}


%
\IEEEpeerreviewmaketitle


\section{Introduction}

\IEEEPARstart{T}{he} evolution of grid management and control, has significantly progressed with the implementation of AMI. AMI has brought many advantages to grid infrastructure and transforms traditional power grids into secure, reliable, and efficient smart grids \cite{ababneh2022private}. According to the National Institute of Standards and Technology (NIST), the smart grid comprises seven domains. \textbf{(i)} bulk generation, \textbf{(ii)} energy transmission, \textbf{(iii)} energy distribution, \textbf{(iv)} customers, \textbf{(v)} market, \textbf{(vi)} operation, and \textbf{(vii)} service provider \cite{gopstein2021nist}. Due to bi-directional communication capability \cite{li2023fine}, domains can seamlessly communicate and interact. This capability enables energy suppliers to offer a wide variety of utilities to their customers, significantly enhancing overall customer service compared to traditional power grids. According to the Juniper research \cite{Juniper2025}, smart grids are projected to save up to \$290 billion in global energy costs by 2029. This demonstrates, how smart grids can significantly impact global energy efficiency and cost saving.

AMI facilitates the integration of renewable energy resources, enabling customers to produce electricity. This has brought financial incentives to consumers, thus they can sell excess energy back to the grid. Additionally, utility providers can benefit from renewable energy resources to generate electricity and substitute fossil fuels with clean alternatives, thereby reducing carbon emissions.

Smart meters play a pivotal role in AMI. They report fine-grained consumption values at predefined intervals (e.g., every 15 or 30 minutes) to utility providers \cite{ibrahem2021privacy}. Such valuable data can be utilized for different purposes such as billing, operation, and value-added services \cite{asghar2017smart}. As of the end of March 2026, there were over 41 million smart and advanced meters operating in homes and small businesses across Great Britain, accounting for 72\% of all meters \cite{DESNZ2026}. In 2024, U.S. electric utilities had installed about 140.5 million AMI meters, approximately 87.6\% of which were residential \cite{EIA2024}. Smart meters not only can reduce customers' costs and save millions of dollars for energy sectors but also can help minimize energy wastage.

Greater granularity and higher frequency of consumption values can bring benefits to data-driven models. These models can be employed to accurately forecast grid loads, detect energy theft, and identify types of appliances in non-intrusive load monitoring applications.  Theoretically, 1 million smart meters with half-hourly reporting can produce about 1460 terabytes of consumption data annually \cite{mitra2024smart}, which is beneficial for neural network models. In the U.S. alone, there are about 135 million utility smart meters, each producing up to 400 megabytes of data annually, resulting in a total of 54 petabytes of new utility data produced each year \cite{HeavyAI2025}. 


In spite of the aforementioned advantages, fine-grained consumption values can reveal consumers' life patterns. An adversary can deduce information about individuals' daily habits, entertainment, sleep schedules, presence or absence at home, and other activities \cite{souri2014smart}. According to this, an adversary with such data granularity can easily violate consumers' privacy. Thus to utilize the full potential of smart grids, we are required to devise privacy-preserving algorithms to address privacy challenges. To enhance customer privacy, various privacy-enhancing protocols have been introduced. Prior works have adopted various approaches such as homomorphic encryption, differential privacy, zero-knowledge proof, and commitment schemes, among others.
While these methods offer significant benefits, they also come with challenges including algorithmic time complexity, computational and communication overhead, high memory requirements, and data utility degradation. Our protocol enables utility providers to collect consumer readings at various levels of granularity while protecting their privacy.

On the other hand, as we mentioned earlier, utility providers tend to offer diverse data-driven value-added services to attract more customers and improve grid management, load monitoring, load forecasting,  infrastructure equipment maintenance, etc. To address these requirements, we propose a \textbf{L}ight \textbf{W}eight \textbf{I}ncentive-Based \textbf{P}rivacy-preserving \textbf{S}mart \textbf{M}etering protocol (LWIPSM) for value-added services. Based on our protocol, utility providers are able to collect consumers' readings in various granularity while protecting customers' privacy and identity. 

Our proposed protocol enables participating smart meters to report coarse-grained or noisy coarse-grained consumption values anonymously while receiving incentives for their participation. Hash-chain credentials and blind digital signatures support anonymous source authentication and issuance--reporting unlinkability, while temporal-based aggregation, local differential privacy, pseudonyms, and anonymous overlay networks protect consumption-value privacy and participant identity. Consequently, the utility provider can collect privacy-preserving consumption data with adjustable granularity and noise levels and construct valuable datasets for data-driven services. \\
 
The main contributions of our proposed protocol are summarized as follows:
\begin{enumerate}
	\item We introduce privacy-preserving incentive programs that allow the utility provider to specify the reporting granularity, program duration, reward amount, data collection purpose, and noise scale. This design enables customers to select a program according to their privacy preferences while allowing the utility provider to collect data with different levels of granularity and utility.
	
	\item We employ hash-chain credentials together with blind digital signatures to provide anonymous source authentication for participating smart meters. The hash-chain credentials enable efficient verification of successive reports, while the blind-signature mechanism separates authenticated program participation from subsequent anonymous reporting. Consequently, the utility provider can verify authorized participation without linking the reported credentials to the meter's real identity, thereby supporting issuance--reporting unlinkability.
	
	\item We combine temporal-based aggregation, local differential privacy, pseudonyms, and an anonymous overlay network to protect both consumption values and participant identities during data collection. Temporal aggregation reduces the granularity of reported readings, while local differential privacy provides a formal privacy guarantee when noise is enabled. Pseudonyms and anonymous communication further conceal the reporting meter from the semi-trusted aggregator and utility provider. This combination enables the collection of privacy-preserving datasets with adjustable granularity and noise levels while maintaining their utility for value-added services.
\end{enumerate}

The remainder of the paper is organized as follows. Section II discusses related work. Section III presents the system model, threat model, design goals, and assumptions. Section IV gives the necessary preliminaries. In Section V, we describe our proposed lightweight incentive-based privacy-preserving smart metering protocol. Section VI evaluates our scheme. Finally, Section VII concludes the paper.


\section{Related Work}

Smart-grid infrastructure supports a wide range of value-added services (VASs) based on smart-meter data. For the purpose of this work, we organize the most relevant privacy-preserving approaches into three functional groups: (1) demand response and energy trading, (2) anonymous reporting and rewarding, and (3) data-driven services. These groups are not necessarily disjoint; for example, anonymously collected readings can subsequently be utilized for data-driven services. Our work lies at the intersection of anonymous incentivized reporting and privacy-preserving data collection for such services.

\subsection{Demand Response and Trading Services}

Privacy-preserving demand response and energy-trading schemes mainly protect customers' consumption profiles, bids, or rewards. Abidin et al. \cite{abidin2016mpc,abidin2018secure} use secret sharing and secure multiparty computation for decentralized electricity trading, although scalability becomes challenging as the number of bids increases. Xue et al. \cite{xue2018ppso} employ Paillier encryption and aggregation for real-time pricing, but assume a fully trusted utility provider. Hassan et al. \cite{hassan2022differentially} instead apply differential privacy to incentivized dynamic pricing, perturbing fine-grained readings before their use in billing and peak-period detection. More recently, Ahmed et al. \cite{ahmed2026p2p} proposed a privacy-preserving dynamic pricing framework for peer-to-peer energy trading that applies local differential privacy while considering incentive-based demand reduction and grid-data utility.

Several incentive-based schemes provide stronger identity protection through cryptographic protocols. Gong et al. \cite{gong2015privacy}, Balli et al. \cite{balli2017distributed}, and Rahman et al. \cite{rahman2017privacy} employ combinations of identity-committable and partially blind signatures, ElGamal encryption, Schnorr signatures, and zero-knowledge proofs to protect bids or demand-response participation. These schemes provide anonymity or unlinkability properties, but zero-knowledge proofs introduce additional computational overhead, while Gong et al. also rely on semi-trusted proxies that raise collusion concerns. Paverd et al. \cite{paverd2014privacy,paverd2014security} avoid some of these cryptographic costs by using a trusted anonymizing entity; however, this entity becomes a single point of trust.

\subsection{Anonymous Reporting and Rewarding Services}

Anonymous reporting schemes are more closely related to our objective because they protect participant identity while supporting rewards. Dimitriou et al. \cite{dimitriou2015enabling} combine blind signatures, hash chains, and zero-knowledge proofs to support anonymous reporting and reward redemption, including double-spending protection. Their later REWARDS framework \cite{dimitriou2019rewards} improves unlinkable reward accumulation and redemption, but primarily protects the rewarding process rather than providing adjustable privacy for the submitted consumption values. More recently, Pei et al. \cite{pei2024blockchain} combine ECC-based anonymous authentication with blockchain-based AMI data aggregation. Their scheme protects identities and aggregated readings, but does not provide incentive-based collection with adjustable reporting granularity and local perturbation.

\subsection{Data-Driven Services}

Privacy-preserving data-driven approaches increasingly avoid centralized disclosure of raw smart-meter readings. Zhang et al. \cite{zhang2020privacy} use differential privacy, data minimization, and local computation for several VASs, although the privacy of the initial training dataset is not clearly addressed. Their more recent federated-learning approach \cite{zhang2024vasfl} keeps consumption data local and exchanges differentially private model information instead, thereby supporting privacy-preserving model training rather than construction of a reusable centralized dataset.

Other works protect specific analytical services through secure or encrypted computation. PrivGrid \cite{lei2024privgrid} securely performs load-data collection, clustering, training, and individual load forecasting without revealing customers' historical readings. Ibrahem and Fouda \cite{ibrahem2024lightweight} use functional encryption for privacy-preserving forecasting, monitoring, and dynamic billing, while He et al. \cite{he2026hierarchicalfl} combine hierarchical federated learning with functional encryption for residential load forecasting. These approaches provide strong protection for predefined computations, but their objective is secure execution of specific services rather than anonymous collection of a reusable privatized dataset.

Distributed-learning and synthetic-data approaches provide another direction. Federated-WDCGAN \cite{chen2023federatedwdcgan} generates synthetic consumption profiles without sharing original datasets, while DP$^2$-NILM \cite{dai2024dp2nilm} and pNILM \cite{wu2025pnilm} protect non-intrusive load monitoring through federated learning, differential privacy, or encrypted computation. In these schemes, raw readings either remain distributed or are replaced by synthetic data. DPP-GAN \cite{li2025dppgan} is particularly relevant because it combines privacy-preserving smart-meter data generation with an incentive mechanism; however, the incentives reward contributions to federated model training and the resulting data are synthetic rather than anonymously authenticated reports of actual consumption values.

Overall, existing approaches protect smart-meter data through cryptographic protocols, anonymous authentication, differential privacy, secure computation, federated learning, or synthetic-data generation. However, among the representative approaches reviewed above, we do not identify a scheme that simultaneously supports incentive-based participation, anonymous authentication of successive actual meter reports, adjustable reporting granularity and local noise, and construction of a reusable privacy-preserving dataset at the utility provider. Our protocol targets this combination by enabling customers to anonymously report coarse-grained or locally perturbed consumption values under selectable privacy-preserving incentive programs.

Table~\ref{tab:related-work-comparison} provides a focused comparison with a subset of representative and closely related schemes selected from the broader literature reviewed above. The table emphasizes the works that are most relevant to our design objectives, particularly anonymous reporting, incentive mechanisms, privacy-preserving use of actual smart-meter readings, adjustable privacy parameters, and reusable data collection.

\begin{table*}[!t]
	\centering
	\caption{Comparison of representative privacy-preserving smart-grid value-added service schemes.}
	\label{tab:related-work-comparison}
	\scriptsize
	\setlength{\tabcolsep}{4.0pt}
	\renewcommand{\arraystretch}{1.15}
	
	\begin{tabularx}{\textwidth}{
			@{}
			>{\raggedright\arraybackslash}p{2.6cm}
			>{\centering\arraybackslash}p{1.45cm}
			>{\centering\arraybackslash}p{1.25cm}
			>{\centering\arraybackslash}p{1.55cm}
			>{\centering\arraybackslash}p{1.75cm}
			>{\centering\arraybackslash}p{1.65cm}
			>{\raggedright\arraybackslash}X
			@{}}
		\toprule
		\textbf{Scheme} &
		\textbf{Identity Privacy} &
		\textbf{Incentive} &
		\textbf{Actual Reports Collected} &
		\textbf{Adjustable Granularity / Noise} &
		\textbf{Reusable Dataset} &
		\textbf{Main Distinction} \\
		\midrule
		
		Gong et al. \cite{gong2015privacy} &
		Yes &
		Yes &
		Yes &
		No &
		No &
		Targets incentive-based demand response and reward allocation. \\
		
		Dimitriou et al. \cite{dimitriou2015enabling,dimitriou2019rewards} &
		Yes &
		Yes &
		Yes &
		No &
		No &
		Supports anonymous reporting and privacy-preserving rewarding. \\
		
		Pei et al. \cite{pei2024blockchain} &
		Yes &
		No &
		Yes &
		No &
		No &
		Provides anonymous authentication and secure AMI data aggregation. \\
		
		Zhang et al. \cite{zhang2024vasfl} &
		No &
		No &
		No &
		Noise only &
		No &
		Keeps raw readings local and performs privacy-preserving federated model training. \\
		
		Lei et al. \cite{lei2024privgrid} &
		No &
		No &
		Yes &
		No &
		No &
		Protects load data during privacy-preserving forecasting computations. \\
		
		Ibrahem and Fouda \cite{ibrahem2024lightweight} &
		No &
		No &
		Yes &
		No &
		No &
		Enables predefined computations over protected meter readings using functional encryption. \\
		
		Chen et al. \cite{chen2023federatedwdcgan} &
		No &
		No &
		No &
		No &
		Yes$^{\dagger}$ &
		Generates reusable synthetic consumption profiles instead of collecting actual reports. \\
		
		Li et al. \cite{li2025dppgan} &
		No &
		Yes &
		No &
		No &
		Yes$^{\dagger}$ &
		Rewards federated model contributions and generates synthetic smart-meter data. \\
		
		\midrule
		\textbf{Our scheme} &
		\textbf{Yes} &
		\textbf{Yes} &
		\textbf{Yes} &
		\textbf{Yes} &
		\textbf{Yes} &
		\textbf{Supports anonymous authenticated collection of actual readings with selectable granularity and local noise for subsequent VASs.} \\
		
		\bottomrule
		\multicolumn{7}{@{}l}{\footnotesize $^{\dagger}$Reusable synthetic dataset rather than anonymously collected actual meter reports.}
	\end{tabularx}
\end{table*}

\section{Problem Statement and Design Goals}
The collection of consumers' fine-grained consumption data poses significant privacy risks. This data can potentially be misused by energy suppliers for household activity inference, behavioral profiling, or unauthorized surveillance. However, utility providers require energy usage data to deliver essential services, including dynamic billing, real-time grid management and monitoring, and value-added services. To address privacy challenges and preserve data privacy while maintaining data utility, we propose a lightweight, incentive-based, privacy-preserving smart metering protocol for value-added services.

Our protocol, allows energy suppliers to obtain consumer's power usage data with adjustable granularity and noise levels. Additionally, our scheme conceals customer identity from energy suppliers by using pseudonyms and anonymous overlay networks. Therefore, energy suppliers can collect a valuable datasets for a broad range of data-driven services in the energy sector. The proposed scheme, employs various privacy-preserving approaches and techniques, including local differential privacy, blind digital signature, hash chain functions, data minimization, anonymity through pseudonyms, and anonymous overlay  networks. 

Before outlining the proposed scheme in detail, we first establish the protocol's system model, threat model, design objectives, and underlying assumptions.

\subsection {System Model}
\begin{center}
	\begin{figure}
		\includegraphics[width=3.6in]{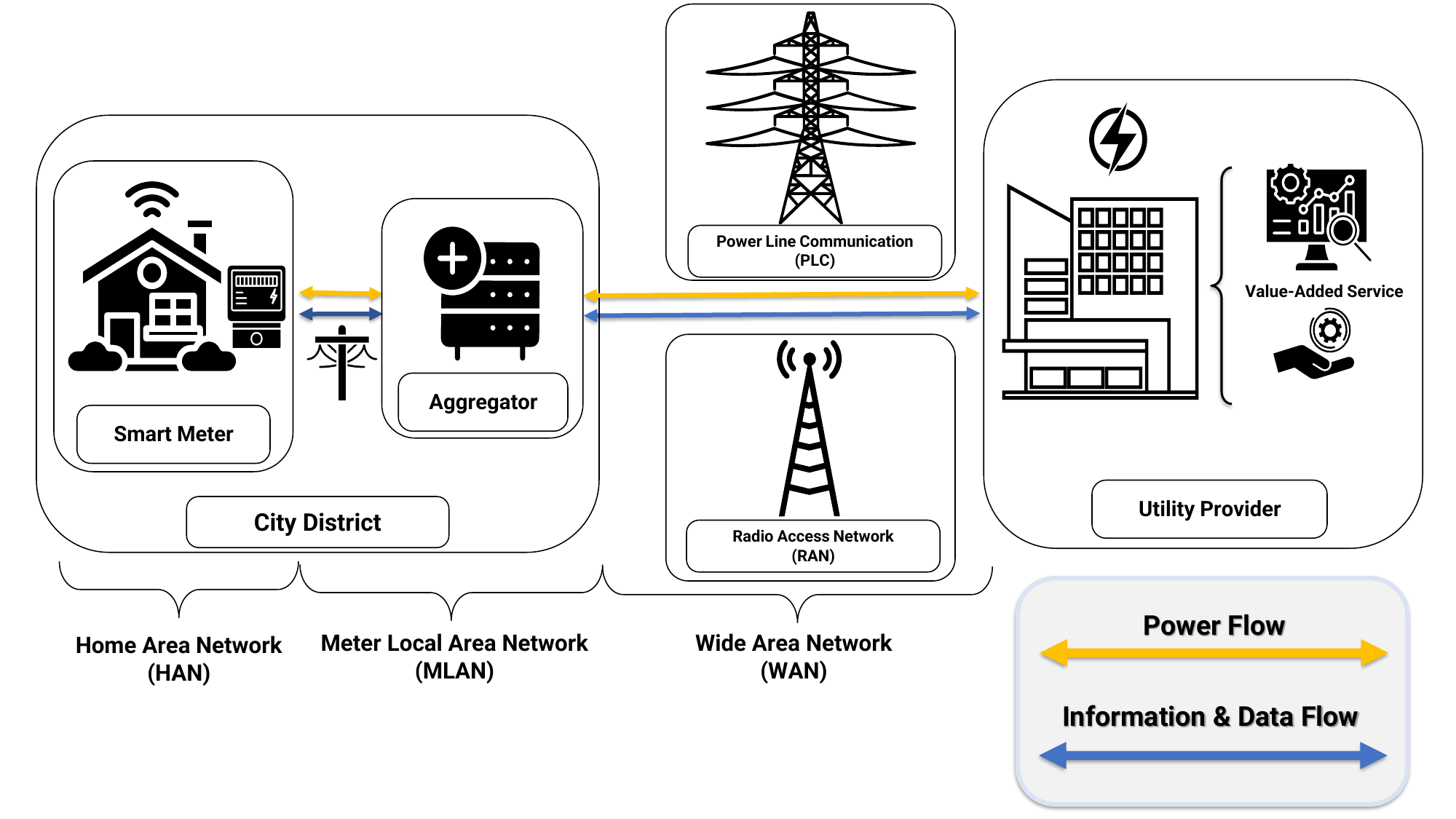}
		\caption{The smart grid's system model consists of smart meters, the aggregator, and the utility provider.}
		\label{vas-system-model}
	\end{figure}
\end{center}

The smart grid's system model contains three entities, including a smart meter, an aggregator, and a utility provider, as shown in Fig. \ref{vas-system-model}.

\begin{enumerate}
	\item \textbf{Smart meter:} Smart meters are intelligent devices measuring consumers' electricity usage, which are connected to the aggregator. Smart meters have limited computational power. They forward consumption data at regular intervals (e.g., every 15 minutes) to the aggregator. Although smart meters have computational constraints, they  perform cryptographic operations such as encryption, decryption, signing, verification, and noise application.
	\item \textbf{Aggregator:}  Aggregators serve as intermediary nodes connecting smart meters to the utility provider. They relay encrypted consumption data to the control center. They can also be utilized as fog/edge nodes to outsource some of the computations. In our scheme, aggregators have sufficient computational resources.
	\item \textbf{Utility Provider:} The utility provider collects consumption values at predefined intervals, stores them, and performs various analyses on collected data. Additionally, it has enough computational resources to offer a wide range of utilities such as billing, operational, and value-added services. 
\end{enumerate}

\subsection {Threat Model and Assumptions}
The protocol defines four key entities categorized into two groups: (1) Legitimate, and (2) Illegitimate entities. Each category is defined as follows: \\
\begin{enumerate}
	\item \textbf{Legitimate Entity:} An entity authorized under grid energy infrastructure (or having consents/agreements) and permitted to participate in the measuring, transmitting, gathering, controlling, monitoring, management, and analysis of smart metering data, as well as offering services. Various levels of trust are defined for each entity within this group.
	\item \textbf{Illegitimate Entity:} Adversaries aiming to launch passive and active attacks on the smart grid infrastructure to gain insight into metering data and violate privacy. Adversaries, also known as untrusted entities, can violate confidentiality, integrity, and privacy.
\end{enumerate}
Smart meters, aggregators, and utility providers are legitimate entities that run and adhere to the protocol. Conversely, active and passive attackers within the network may attempt to eavesdrop, intercept, or even modify metering data during transmission. These malicious activities raise concerns about confidentiality, integrity, and privacy. 

\begin{center}
	\begin{figure}[htp]
		\includegraphics[width=3.6in]{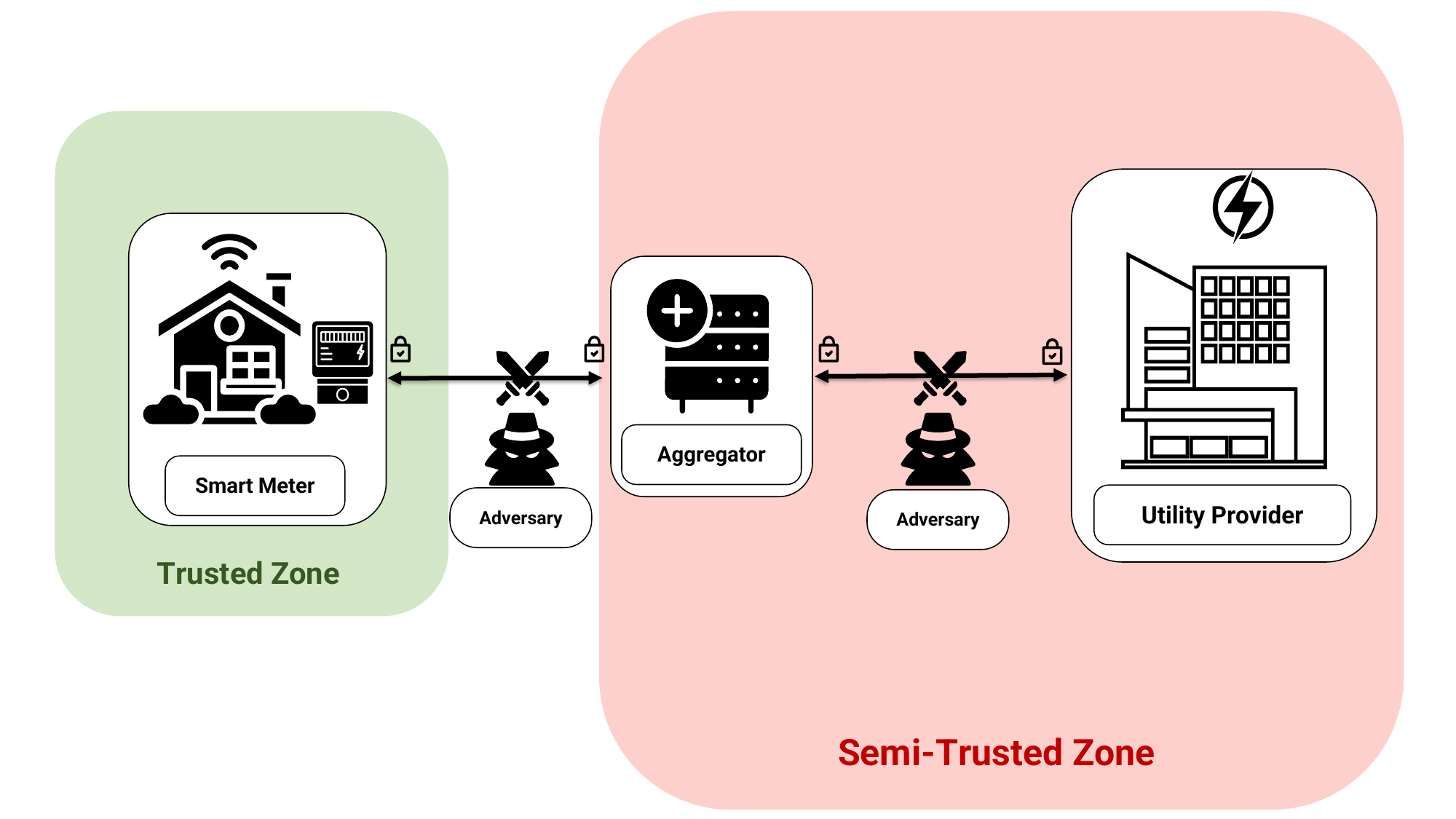}
		\caption{The protocol's threat model, which protects user's consumption data privacy against semi-trusted and untrusted entities.}
		\label{vas-threat-model}
	\end{figure}
\end{center}

In our scheme, each legitimate entity has various levels of trust, therefore we define our threat model as follows:

\begin{enumerate}
	\item \textbf{Smart meters:} Smart meters are trusted and tamper resistant. They are equipped with tamper-proof protection mechanisms, ensuring, that any  attempts by customers to perform reverse engineering or physical attacks (including invasive, non-invasive, and semi-invasive attacks) trigger an alarm and inform the utility provider. Subsequently, the meter's status changes to non-functional and erases its security parameters (e.g., cryptographic keys). We also assume that meters do not collude with other entities or even with each other.
	\item \textbf{Aggregators:} Aggregators are trusted but curious (semi-trusted), which means, they adhere to the protocol but may attempt to invade customers' privacy out of curiosity. Aggregators may collaborate with the utility provider but never collude with adversarial actors.
	\item \textbf{Utility provider:} The utility provider is also semi-trusted. The utility provider may cooperate with third-party service providers under certain privacy regulations. In our scheme, we assume that an additional layer of privacy protection is implemented to safeguard customers' privacy in case of collaboration. The utility provider never collaborates with untrusted entities. 
	\item \textbf{Adversary:} As we mentioned earlier, adversaries attempt to launch passive/active attacks on energy grid networks to compromise confidentiality, integrity, and privacy. They are present only within the semi-trusted boundary and operate solely within this domain.
\end{enumerate}
The components involved in the protocol's threat model are shown in Fig. \ref{vas-threat-model}. As shown in this figure, there are two zones: (1) Trusted, which includes smart meters, and (2) Semi-Trusted, which includes aggregators and the utility provider.

The proposed scheme is based on the following assumptions:
\begin{enumerate}
	\item All smart meters are registered and authenticated before the protocol is initiated.
	\item Legitimate entities securely distribute their public key among each other.
	\item Smart meters are trusted by both customers and the utility provider. They do not collude and always obey the protocol. Meters always act in a trusted manner.
\end{enumerate}

\subsection{Design Goals}
In our proposed incentive-based privacy-preserving protocol, we attempt to achieve the following goals:

\begin{enumerate}
	\item \textbf{Privacy:} Customers' privacy should be protected against semi-trusted and untrusted entities. Semi-trusted entities are not able to map fine-grained consumption values to the corresponding customers, thereby  consumers' load profiles remain concealed. Additionally, privacy is enhanced by adding noise to consumption values using local differential privacy. Moreover, Anonymity and unlinkability are two essential privacy factors that should be ensured.
	\begin{itemize}
		\item \textbf{Anonymity:} Anonymity of a smart meter means that a smart meter is not identifiable within a set of smart meters. referred to as anonymity set \cite{pfitzmann2010terminology}. 
		\item \textbf{Unlinkability:} Unlinkability of two or more items of interest from the perspective of semi-trusted or untrusted entities, means that within the grid system infrastructure, they cannot distinguish sufficiently whether these items of interest are related or not \cite{pfitzmann2010terminology}.
	\end{itemize}
	\item \textbf{Confidentiality:} Our scheme aims to ensure the confidentiality of consumption data against untrusted illegitimate entities. This ensures adversarial actors are not able to deduce any information from intercepted encrypted data. 
	\item \textbf{Integrity:} The scheme guarantees data integrity. Any data alteration or modification attack by illegitimate entities is detected using a message authentication code and digital signature schemes.
	\item \textbf{Utility:} The protocol maintains a balance between privacy and utility, meaning that while preserving privacy, it also ensures data utility for various services. 
	\item \textbf{Feasibility in Smart Grid Infrastructure:} The computational power required to execute the protocol is compatible with the capabilities of the existing devices in the smart grid network.
\end{enumerate}

\section{Preliminaries}

As we mentioned earlier, our protocol incorporate several privacy protection mechanism including local differential privacy, RSA blind digital signatures, and anonymous overlay networks. Before introducing the proposed scheme, the main techniques that are utilized in our scheme are introduced in the rest of this section.

\subsection {Differential Privacy}
Differential Privacy is a mathematical framework that protects data privacy while enabling data analysis. In 2006, Dwork et al. \cite{dwork2006calibrating} introduced the concept of $\epsilon$-differential privacy. Differential privacy is categorized into two trust models:
\begin{enumerate}
	\item \textbf{Global (Central) Differential Privacy:} In this model,  a trusted curator collects original data from data sources (in our case, meters) and applies a privatized mechanism (or randomized algorithm) on original data before publishing it.
	\item \textbf{Local Differential Privacy:} In this model, each individual data source (in our case, meters) perturbs its own data before sending it to the curator. The curator is assumed to be semi-honest or even untrusted.
\end{enumerate}
Now let's delve deeper into the frameworks of local and global differential privacy and formally define each concept.

\subsubsection {Global Differential Privacy}
Let $D$ and $D^{'}$ be two datasets that only differ in one record (i.e., $D$ and $D^{'}$ are neighboring datasets). A privatized mechanism $\mathcal{M}$ satisfies $\epsilon$-differential privacy if for all such neighboring datasets $D$ and $D^{'}$, and for all sets of possible outputs $S$, the following relation holds:
\begin{align}
	Pr[\mathcal{M}(D) \in S] \leq e^{\epsilon} \cdot Pr[\mathcal{M}(D^{'}) \in S] 
	\label{vas-gdp-relation}
\end{align}
In this relation, $\epsilon$ is the privacy budget. A smaller $\epsilon$ indicates stronger privacy.

\subsubsection{Local Differential Privacy} Let $x_i$ and $x_i'$ denote any two possible private inputs of user $i$ in a bounded domain $\mathcal{X}$. A randomized mechanism $\mathscr{M}$ satisfies $(\epsilon,\delta)$-local differential privacy if, for every measurable output set $S\subseteq Range(\mathscr{M})$, \begin{align} \Pr[\mathscr{M}(x_i)\in S] \leq e^{\epsilon}\Pr[\mathscr{M}(x_i')\in S]+\delta . \label{vas-ldp-relation} \end{align} Here, $\epsilon>0$ controls the distinguishability between the two inputs, while $0\leq\delta<1$ bounds the probability of exceptional privacy loss. When $\delta=0$, the definition reduces to pure $\epsilon$-LDP.

\subsection {RSA Blind Digital Signature}
The RSA blind digital signature \cite{chaum1983blind} is a two-party protocol that has the following two properties: 
\begin{enumerate}
	\item \textbf{Blindness:} The signer cannot obtain the original message he/she is signing. 
	\item \textbf{Unlinkability:} After the signature is unblinded, the signer cannot link between the unblinded signature and the blinded message.
\end{enumerate}
This scheme consists, key generation, message blinding, signing blinded message, and blind factor removal phase, which are explained in detail as follows: 
\begin{enumerate}
	\item \textbf{Key Generation:}
	The utility provider generates an RSA blind-signature key pair 	$(pk_{UP}^{\mathrm{bs}},sk_{UP}^{\mathrm{bs}})$, where 	$pk_{UP}^{\mathrm{bs}}=(e,N)$ and $sk_{UP}^{\mathrm{bs}}=(d,N)$. Messages and credentials used in the blind-signature scheme are represented as elements of $\mathbb{Z}_N$.
	\item \textbf{Message Blinding:}
	\begin{itemize}
		\item Choose a random value $r\in\mathbb{Z}_N^{*}$.
		\item Compute the blinded message as $m^{'} = m \cdot r^{e} \bmod N$.
		\item Blinded message $m^{'}$ is sent to the signer.
	\end{itemize}
	\item \textbf{Signing Blinded Message:}
	\begin{itemize}
		\item Signer computes $\sigma_{m^{'}}^{signer}=(m^{'})^d \bmod N$ 
		\item Signer sends back the blinded signature $\sigma_{m^{'}}^{signer}$
	\end{itemize}
	\item \textbf{Blind Factor Removal:}
	\begin{itemize}
		\item Receiver removes the blind factor: \\ 
		$\sigma_{{m}}^{signer} = \sigma_{m^{'}}^{signer} \cdot r^{-1} \bmod N$
	\end{itemize}
\end{enumerate}
In our case, using the RSA blind digital signature scheme, a meter can obtain the utility provider's signature on any arbitrary message $m$.
\subsection{Anonymous Overlay Networks}

In our scheme, smart meters utilize an anonymous overlay network to forward their encrypted readings toward the aggregator and utility provider. An anonymous overlay network is a communication network built on top of an existing network (e.g., the Internet) that aims to conceal the identities of participants or the relationship between senders and receivers. More specifically, an anonymous overlay network provides three types of anonymity:
\begin{itemize}
	\item \textbf{Sender Anonymity:} A message cannot be linked to the sender \cite{pfitzmann2010terminology,ren2016anonymous}.
	\item \textbf{Receiver Anonymity:} A message cannot be linked to the recipient \cite{pfitzmann2010terminology,ren2016anonymous}.
	\item \textbf{Relationship Anonymity:} The relationship between the sender and receiver cannot be determined or identified \cite{pfitzmann2010terminology,ren2016anonymous}.
\end{itemize}

Our protocol primarily relies on sender anonymity to conceal the source of reported consumption readings. Smart meters forward encrypted reports through neighboring meters in a unidirectional anonymous overlay network until they reach the aggregator. Since the communication is unidirectional, the utility provider cannot directly acknowledge packet reception to the originating meter. To support reliable delivery, a Bloom filter is used to record the status of received packets and is subsequently broadcast to the network, enabling meters to verify whether their reports have been successfully received.

An optimized anonymous overlay network was proposed by Finster et al. \cite{finster2013pseudonymous}, which satisfies the communication and anonymity requirements of our protocol. Accordingly, we adopt Finster's approach for establishing the anonymous overlay network.


\section{Proposed Scheme}
To tackle privacy challenges associated with smart metering data while offering numerous types of utilities such as value-added services to consumers, we propose a privacy-enhancing protocol. This protocol allows energy suppliers to collect consumption readings with adjustable granularity and accuracy in a privacy-preserving manner, facilitating the generation of a valuable large-scale privatized dataset. This dataset can later be utilized by control centers to offer a broad range of value-added services to customers. This section begins with a high-level overview of the scheme to provide an initial insight, followed by a detailed description of the protocol.

\subsection{Protocol Overview}
This protocol contains several steps, which are explained below: 
\begin{enumerate}
	\item At the beginning of the protocol, the utility provider generates a list of privacy-preserving incentive programs. This list specifies various parameters including the granularity (or frequency) of readings, the reward value (token value), program duration, the purpose of data collection, and noise scale.
	
	\item Customers select a privacy-preserving incentive program based on their needs and preferences. Subsequently, the meter creates its own credentials using a collision-resistant hash chain function. These credentials are used for anonymous reporting and source authentication. Additionally, the meter adjusts its value-added service channel frequency to match the transmission frequency of the chosen privacy-preserving incentive program.
	
	\item Customers can choose whether to add local noise to their readings. If a customer chooses a program that enables the meter to add local noise, the meter uses a local differential privacy mechanism to perturb readings within an acceptable privacy budget.
	
	\item The meter blinds the last created credential using a blind factor. Afterward, the blinded credential is signed by the meter. The blinded credential, the chosen program, and the signature are then encrypted and sent to the utility provider.
	
	\item After decrypting and successfully verifying the meter's request, the utility provider generates and signs the token corresponding to the chosen program and blind-signs the meter's blinded credential. It then signs and encrypts the generated token, token signature, and blinded-credential signature, and sends the resulting message back to the meter.
	
	\item The meter first verifies the received message. Upon successful verification, it removes the blind factor from the blinded credential signature to obtain the utility provider's signature on the last credential.
	
	\item The meter generates a pseudonym for anonymous reporting and substitutes its original identity with the pseudonym. It then prepares the first reporting message containing the reporting epoch and sequence number, the last credential, the utility provider's signature on the last credential, the first coarse-grained consumption value (following a temporal-based aggregation scheme), and the pseudonym. The meter computes a MAC over the required reporting fields using the program-specific MAC key $k_s^{\mathrm{MAC}}$ and encrypts the resulting message using the utility provider's public encryption key.
	
	\item The prepared encrypted message in the previous step is transmitted to the utility provider through a unidirectional anonymous overlay network among neighboring meters.
	
	\item The utility provider first decrypts the received message and verifies the last credential using the credential, the utility provider's signature on the credential, and its public key. Upon successful verification, it stores the last credential for the corresponding pseudonym as the latest accepted credential of the anonymous meter.
	
	\item The utility provider verifies the integrity of the required reporting fields, including the reporting epoch, sequence number, coarse-grained consumption value, and pseudonym, using the program-specific MAC key $k_s^{\mathrm{MAC}}$. Upon successful verification, it stores the consumption value.
	
	\item For each subsequently acknowledged report, the meter advances to the next credential, which is verified against the latest accepted credential using the hash-chain relation. After successfully accepting a report, the utility provider includes its acknowledgment identifier in a signed Bloom filter corresponding to the reporting slot and broadcasts it through the aggregator. The meter advances to the next report only after observing a positive acknowledgment; otherwise, it retransmits the same report after a predefined timeout. After the final report is successfully acknowledged, the token becomes available for redemption according to its predefined activation delay.
\end{enumerate}

According to the protocol, the utility provider is able to collect meter readings in privacy-preserving manner to create a dataset. This dataset can then be used for data-driven services such as machine learning, deep learning, and data mining or other purposes. 

\begin{center}
	\begin{figure}[htp]
		\includegraphics[width=3.5in]{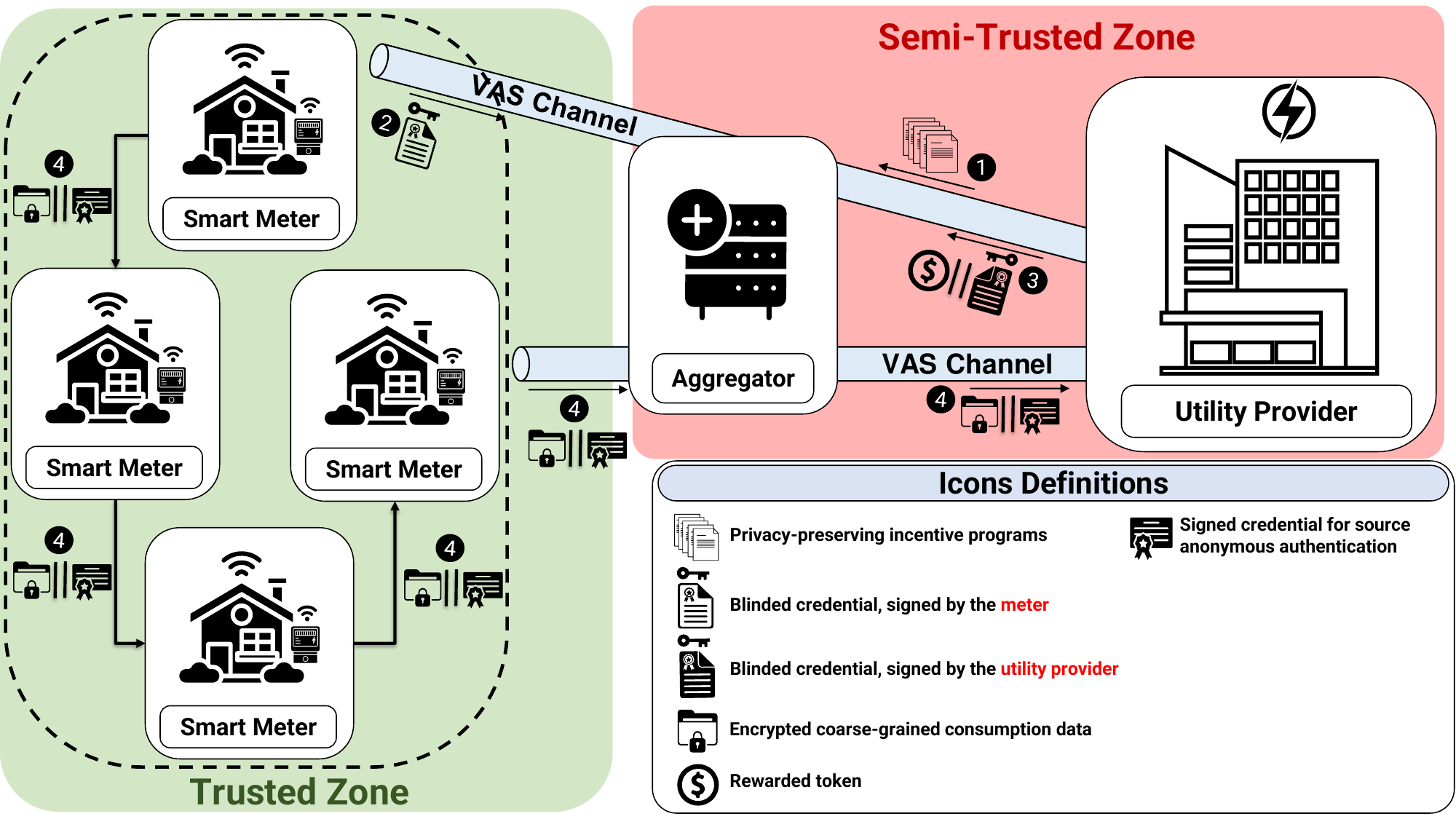}
		\caption{Our proposed lightweight incentive-based privacy-preserving smart metering protocol for value-added services.}
		\label{vas-protocol-overview}
	\end{figure}
\end{center}

As we mentioned earlier, meters are trusted and never deviate from the protocol. According to this, it is important to note that the token is stored by the meter, and the customer has no access to it until the meter grants permission. 
The protocol summary is depicted in Fig. \ref{vas-protocol-overview}.

\subsection{Protocol Details}
In this section, we examine the details of the proposed scheme. The main symbols employed in the specification of the protocol are summarized in Table \ref{table:vas-protocol-elements}.


\begin{table}[htp]
	\centering
	\caption{Protocol Notations}
	\renewcommand{\arraystretch}{1.5} 
	\setlength{\tabcolsep}{5pt} 
	\resizebox{1.006\columnwidth}{!}{ 
		\begin{tabular}{|p{3cm}|p{4cm}|p{6.9cm}|}
			\hline
			\textbf{Symbols} & \textbf{Description} & \textbf{Formal Representation} \\
			\hline
			\( M \) & Number of registered and active meters in the network & \( M \in \mathbb{N} \) \\
			\hline
			\( PR \) & List of privacy-preserving incentive programs & \( PR = \langle pr_1, pr_2, \ldots, pr_g \rangle \) \\
			\hline 
			\( pr_i \) & A privacy-preserving incentive program that includes a unique program identifier, frequency, token information, program duration, program activation time, data collection purpose, and noise-scale parameter & \( pr_i = \langle pid_i, freq_i, tokinf_i, pd_i, pat_i, prp_i, nsc_i \rangle \) \\
			\hline
			\( \mathscr{N} \) & Number of participating customers in a specific program (e.g., \(pr_i\)) & \( \mathscr{N} \in \mathbb{N} \) \\
			\hline
			\( pid_i \) & Unique identifier of the \(i\)th privacy-preserving incentive program & \( pid_i \in \{0,1\}^{\lambda} \) \\
			\hline
			\( freq_i \) & \( i \)th reporting frequency & \( freq_i \in FREQ \) \\
			\hline
			\( tokinf_{i} \) & \( i \)th token information details: value, validity period, activation delay & \( tokinf_{i} = \langle value, validDays, tokActivDelay \rangle \in TOKINF \) \\
			\hline
			\( tok_i \) & \( i \)th token details: value, expiration, activation time, a unique ID & \( tok_i = \langle value, expDateTime, ActivDateTime, uid_i \rangle \in TOKEN \) \\
			\hline
			\( pd_i \) & \( i \)th program duration & \( pd_i \in PD \) \\
			\hline
			\( pat_i \) & Activation date-time of privacy-preserving incentive program & \( pat_i \in \mathbb{\text{DateTime}} \) \\
			\hline
			\( prp_i \) & \( i \)th purpose for data collection & \( prp_i \in PRP \) \\
			\hline
			\( nsc_i \) & \( i \)th noise scale parameter for data perturbation & \( nsc_i \in NSC \) \\
			\hline
			\( FREQ \) & Set of reporting frequencies & \( FREQ = \{freq_1, freq_2, \ldots , freq_x\} \) \\
			\hline
			\( TOKINF \) & Set of defined token information (informational) & \( TOKINF = \{tokinf_1, tokinf_2, \ldots, tokinf_y\} \) \\
			\hline
			\( TOKEN \) & Set of defined tokens (spendable)& \( TOKEN = \{tok_1, tok_2, \ldots, tok_y\} \) \\
			\hline
			\( PD \) & Set of program durations & \( PD = \{pd_1, pd_2, \ldots, pd_z\} \) \\
			\hline
			\( PRP \) & Set of purposes for data collection & \( PRP = \{prp_1, prp_2, \ldots, prp_k\} \) \\
			\hline
			\( NSC \) & Set of noise scale parameters for data perturbation & \( NSC = \{nsc_1, nsc_2, \ldots, nsc_l\} \) \\
			\hline
			\( V_{id_j} \) & Initial random value from smart meter \( id_j \) & \( V_{id_j}=cr_{0, id_j}\in \mathbb{CR}_{id_j} \) \\
			\hline
			\( cr_{n, id_j} \) & \( n \)th certificate of meter \( id_j \) & \( cr_{n, id_j} \in \mathbb{CR}_{id_j} \)\\
			\hline
			\( \sigma_{cr_{n, id_j}}^{UP} \) & Service provider’s signature on \( n \)th certificate & \( \sigma_{cr_{n, id_j}}^{UP} \in \mathbb{\sigma}_{cr_n}^{UP} \)\\
			\hline
			\( uuid_{\omega} \) & Pseudonym generated by meter & \( uuid_{\omega} \in UUID \) \\
			\hline
			\( t_i \) & \( i \)th time interval  & \( T = \langle t_1 , t_2, \ldots, t_i , \ldots, t_n \rangle \) \\
			\hline
			\( c_{t_i, uuid_{\omega}}\) & consumption value of a meter at \( i \)th interval with pseudonym of \( uuid_{\omega} \) & \( c_{t_i, uuid_{\omega}} \in \mathbb{C}_{uuid_{\omega}} \) \\
			\hline
			\( H \) & A preimage-resistant and collision-resistant hash function for credential generation & \( H: \{0,1\}^{*}\rightarrow\{0,1\}^{\ell} \) \\
			\hline
			\( bf_{id_j} \) & The meter's random RSA blinding factor & \( bf_{id_j} \in \mathbb{Z}_N^{*} \) \\
			\hline
			\( sk_{id_j}, pk_{id_j} \) & Private and public key pairs of meter \( id_j \) & \( sk_{id_j} \in \mathbb{SK}, pk_{id_j} \in \mathbb{PK} \) \\
			\hline
			\( sk_{AGG}, pk_{AGG} \)  & Private and public key pairs of aggregator & \( sk_{AGG} \in \mathbb{SK}, pk_{AGG} \in \mathbb{PK} \) \\
			\hline 
			\( (sk_{UP}^{x},pk_{UP}^{x}),\;x\in\{\mathrm{enc},\mathrm{sig},\mathrm{bs}\} \) & Independent key pairs of the utility provider for encryption,
			ordinary digital signatures, and RSA blind signatures & \( sk_{UP}^{x}\in\mathbb{SK},\; pk_{UP}^{x}\in\mathbb{PK} \) \\
			\hline
			\( k_{s}^{\mathrm{MAC}} \) & A fresh program-specific MAC key for program \(pr_s\), shared among the utility provider, the aggregator, and the smart meters participating in \(pr_s\) for data integrity checking & \( k_{s}^{\mathrm{MAC}} \in \mathbb{SKEY}\) \\
			\hline
			 \( || \)  & This symbol represents the concatenation of messages, such as $m_1$ and $m_2$ & \( m_1 || m_2\) \\
			\hline
		\end{tabular}
	}
	\label{table:vas-protocol-elements}
\end{table}
\vspace{2mm}
Throughout the protocol, $Enc(pk,\cdot)$ and $Dec(sk,\cdot)$ denote secure hybrid encryption and decryption operations, respectively. For each payload, a fresh symmetric session key is used to encrypt the payload, while the session key is protected using the recipient's public key. For notational simplicity, we abstract the complete hybrid encryption process, including both the protected session key and the encrypted payload, by the single notation $Enc(pk,M)$. Accordingly, the protocol is not restricted by the plaintext-size limitation of direct asymmetric encryption. 

\begin{normalsize}
	\noindent \textbf{Phase I - Privacy Incentive Programs Generation:}
\end{normalsize}
This phase aims to generate a list of privacy incentive programs that contain detailed program information, including the program duration, the transmission frequency rate, the token amount, the data collection purpose, and the applicable noise scale. This allows customers to select the appropriate program based on their preferences.

At the beginning of the protocol, the utility provider forwards a list of privacy-preserving incentive programs to a participating smart meter. The customer examines the privacy-preserving incentive programs using an in-home display  and selects a program based on his/her needs. The list $PR$ consists of various programs such as $pr_i$ which is defined as follows.
\begin{IEEEeqnarray}{rCl}
	PR &=& \{pr_1, pr_2, \ldots, pr_g\} \notag \\
	pr_i &=& \langle  pid_i, freq_i, tokinf_{i}, pd_i, pat_i, prp_i, nsc_i \rangle 
	\label{eq:vas-protocol-details-privacy-programs-list}
\end{IEEEeqnarray}
Where $pid_i$ denotes the unique identifier generated by the utility provider for program $pr_i$, $freq_i \in FREQ$ denotes the number of reports per day, $tokinf_{i}=\langle value, validDays, tokActivDelay \rangle$ signifies the corresponding token information for participation in the incentive program $pr_i$, $pd_i \in PD$ represents the program duration, $pat_i \in DateTime$ indicates the activation time of $pr_i$, $prp_i \in PRP$ denotes the data collection purpose, and $nsc_i \in NSC$ defines the program-specific noise-scale parameter used for data perturbation. The utility provider ensures that each generated $pid_i$ is unique among the available programs. The parameters $\epsilon$ and $\delta$ are predefined system-wide baseline privacy parameters common to all programs, while $nsc_i$ adjusts the noise level and determines the effective per-report privacy budget.

An example of a privacy-preserving incentive program $pr_i$, is defined as follows:
\begin{IEEEeqnarray}{rCl}
	pr_i &=& \langle pid_i, 12, \langle 15, 45, \text{"24h"} \rangle, 7, \text{"00:00:00 - 1d"}, \notag \\
	&& \text{"Data-Driven Services"}, 5 \rangle
	\label{eq:vas-protocol-details-privacy-programs-list-example}
\end{IEEEeqnarray}
In this example, $pid_i$ represents the unique randomly generated identifier assigned to the program. $12$ indicates that a smart meter should send 12 reports per day (i.e., every 2 hours). $\langle 15, 45, \text{"24h"} \rangle$ signifies that the utility provider will provide a reward (or token) of \$15, which remains valid for 45 days and activates 24 hours after the last report. $7$ specifies that the program lasts for 7 days. $\text{"00:00:00 - 1d"}$ denotes that the chosen program will start the next day at "00:00:00". $\text{"Data-Driven Services"}$ signifies that the data is used for data-driven services (e.g., as a training set for a deep learning model). $5$ represents the noise scale; customer may choose another program with a noise scale of zero. 

Privacy-preserving incentive programs are constrained to specific frequencies and program duration as defined in Relation \ref{eq:vas-protocol-details-privacy-programs-frequency-programDuration}. By reducing reporting frequency and limiting program duration, we mitigate privacy risks by decreasing the granularity and overall volume of collected data.
\begin{IEEEeqnarray}{rCl}
	freq_i \in FREQ &=& \{4, 6, 8, 12, 16\} \notag \\
	pd_i \in PD &=& \{7, 8, \ldots, 21\}
	\label{eq:vas-protocol-details-privacy-programs-frequency-programDuration}
\end{IEEEeqnarray}
In this study, we suggest candidate frequencies from the set $\{4,6,8,12,16\}$ to provide coarser-grained consumption data compared to standard baselines---such as 144, 96, or 48 reports per day---while maintaining sufficient full-day coverage for a broad range of value-added services and ensuring mathematical consistency to avoid imbalanced sampling. Additionally, selecting a duration of 1 to 3 weeks strikes a balance between collecting sufficient data for meaningful analysis and limiting exposure to privacy risks.

Temporal aggregation enhances customers' privacy by reducing data granularity and consequently, decreases the ability of semi-trusted entities to infer information from anonymous coarse-grained consumption values. Buescher et al. \cite{buescher2017two} demonstrated that increasing temporal resolution reduces the advantage of semi-trusted entities attempting to violate customer privacy. Limiting the program duration to 1 to 3 weeks ensures that fewer data points are collected, minimizing the impact of potential breaches. At the same time, given a sufficient number of participating meters in each privacy program, the utility provider can still gather an adequate amount of data essential for various purposes (e.g., model training or energy optimization).

As the utility provider generates a list of privacy-preserving incentive programs, it utilizes a reward calculation method for computing $tokinf_i$, which consists of three elements: $value_i$, $validDays_i$, and $tokActivDelay_i$. Here, $value_i$ represents the token amount, $validDays_i$ specifies the number of days for which the token remains valid, and $tokActivDelay_i$ denotes the token activation delay after the last report. The reward calculation method is defined as follows:
\begin{IEEEeqnarray}{rCl}
	value_i &=& B_{\mathrm{val}}
	+ (w_{\mathrm{freq}}^{\mathrm{val}} \cdot freq_i)
	+ (w_{\mathrm{pd}}^{\mathrm{val}} \cdot pd_i) \nonumber\\
	& & + w_{\mathrm{prp}}^{\mathrm{val}}(prp_i)
	- (w_{\mathrm{nsc}}^{\mathrm{val}} \cdot nsc_i)
	\IEEEyesnumber\IEEEyessubnumber
	\label{eq:vas-protocol-details-value-spec} \\[6pt]
	validDays_i &=& B_{\mathrm{dur}}
	+ (w_{\mathrm{freq}}^{\mathrm{dur}} \cdot freq_i)
	+ (w_{\mathrm{pd}}^{\mathrm{dur}} \cdot pd_i) \nonumber\\
	& & + w_{\mathrm{prp}}^{\mathrm{dur}}(prp_i)
	- (w_{\mathrm{nsc}}^{\mathrm{dur}} \cdot nsc_i)
	\IEEEyessubnumber
	\label{eq:vas-protocol-details-validDays-spec}
\end{IEEEeqnarray}
The third element, $tokActivDelay_i$, is set to a specific value  (e.g., $\text{"24h"}$), indicating that the token is activated 24 hours after the last report. Here, $B_{\mathrm{val}}$ and $B_{\mathrm{dur}}$ denote the baseline token value and validity duration, respectively. For each weighting coefficient $w_x^y$, the subscript $x\in\{\mathrm{freq},\mathrm{pd},\mathrm{prp},\mathrm{nsc}\}$ denotes the corresponding program parameter, namely reporting frequency, program duration, data collection purpose, and noise scale, while the superscript $y\in\{\mathrm{val},\mathrm{dur}\}$ indicates whether the weight affects the token value or its validity duration. The utility provider determines these weighting coefficients according to its established incentive policies. Higher reporting frequency and longer program duration increase the token value and validity duration, whereas a higher noise scale decreases them due to the corresponding reduction in data utility. The purpose-dependent terms $w_{\mathrm{prp}}^{\mathrm{val}}(prp_i)$ and $w_{\mathrm{prp}}^{\mathrm{dur}}(prp_i)$ are assigned according to the selected data collection purpose.

It is also important to note that the number of participating smart meters in each program $pr_i$ must exceed a predefined threshold; otherwise, the utility provider may be able to identify the meter's identity. For instance, if only one user participates in a privacy-preserving incentive program, the utility provider can easily detect the meter and link its pseudonym to the real identity. Consequently, the utility provider identifies the user's identity. To prevent this, the protocol enforces the following condition: if the number of participants in a given program does not exceed the predefined threshold, the utility provider cancels the program and notifies the participating customers. This constraint for each program $pr_i$ with $k$ number of participants is defined formally as follows where $ID_{pr_i}$ is the set of smart meters' IDs participating in program $pr_i$.
\begin{IEEEeqnarray}{rCl}
	ID_{pr_i} &=& \{id_1^{(i)}, id_2^{(i)}, \ldots, id_{k}^{(i)} \}, \nonumber\\
	& & \quad \text{if } |ID_{pr_i}| > \text{threshold} \Rightarrow \text{program is executed} \notag \\
	\IEEEyesnumber \label{eq:vas-protocol-details-anonymity-setsize}
\end{IEEEeqnarray}

\begin{normalsize} 
	\noindent
	\textbf{Phase II - Credential Generation and Blinding:}
\end{normalsize}
In this phase, Each meter generates a set of credentials based on the selected program. These credentials serve both as proof of anonymous source authentication and as an indication of the meter’s participation in the program. Additionally, to obtain the utility provider's signature on the last credential, each participating meter blinds the last credential and sends it to the control center for blind digital signing. Since smart meters are trusted and always follow the protocol, each generated credential chain is only used for the selected privacy program and is not reused in other programs.

When a customer selects the privacy program $pr_{s}$, smart meter (with $id_j$) starts to generate an initial random value $V_{id_j}$ as initial credential $cr_{0, id_j}$, then it generates a set of credentials using a secure hash function. The credential generation process is demonstrated as follows:
\begin{IEEEeqnarray}{rCl}
n &=& freq_s \cdot pd_s \notag\\[2pt]
cr_{i, id_{j}} &\leftarrow& H(cr_{i-1, id_j}), i \in [1, n-1] \label{eq:vas-protocol-details-credential-generation}
\end{IEEEeqnarray}
The parameter $n$ indicates the number of credentials during a privacy program. Since each new credential is derived from the previous one, the last credential $cr_{n-1, id_j}$ is used for receiving the utility provider's signature.

After credential generation, the smart meter blinds the last credential $cr_{n-1, id_j}$ using blind factor $bf_{id_j}$ and produces the blinded credential $cr_{n-1, id_j}^{'}$ as follows where $bf_{id_j}$ is the blinding factor of this smart meter.
\begin{IEEEeqnarray}{rCl}
	cr_{n-1,id_j}^{'}
	&\leftarrow&
	cr_{n-1,id_j}\cdot bf_{id_j}^{e}\bmod N
	\label{eq:vas-protocol-details-blind-last-credential}
\end{IEEEeqnarray}
Now, the smart meter signs $cr_{n-1,id_j}^{'} || pr_s$ using its private key $sk_{id_j}$ and then encrypts it using the utility provider's encryption public key $pk_{UP}^{\mathrm{enc}}$.
\begin{IEEEeqnarray}{rCl}
	\mathcal{m} &=& cr_{n-1, id_j}^{'} || pr_{s} \notag \\
	\sigma_{\mathcal{m}}^{id_j} &\leftarrow& Sign(sk_{id_j},\mathcal{m}) \notag \\
	\mathcal{x} &\leftarrow& Enc(pk_{UP}^{\mathrm{enc}}, \mathcal{m} || \sigma_{\mathcal{m}}^{id_j}) \notag \\
	\label{eq:vas-protocol-details-cr-pr-signcryption}
\end{IEEEeqnarray}
The smart meter forwards $\mathcal{x}$ to the utility provider via aggregator.

\vspace{1mm}
\begin{normalsize}
	\noindent \textbf{Phase III - The Token Generation and Blind Digital Signing:}
\end{normalsize}
This phase focuses on generating the token corresponding to the selected program. Additionally, the utility provider signs the token using an ordinary digital signature and signs the last blinded credential using the RSA blind-signature scheme.

The utility provider receives $\mathcal{x}$ and decrypts it. It then verifies the message.
\begin{IEEEeqnarray}{rCl}
	\mathcal{m} || \sigma_{\mathcal{m}}^{id_j}  &\leftarrow& Dec(sk_{UP}^{\mathrm{enc}}, \mathcal{x}) \notag \\
	True/False &\leftarrow& Verify(pk_{id_j}, \mathcal{m}, \sigma_{\mathcal{m}}^{id_j})
	\label{eq:vas-protocol-details-x-decryption-verification}
\end{IEEEeqnarray}
Upon successful verification, the utility provider first generates the corresponding token $tok_s=\langle value, expDateTime, activeDateTime, uid_s\rangle$, where $value$ represents the token amount, $expDateTime$ indicates token expiration date-time, $activeDateTime$ signifies the token activation date-time, and $uid_s$ denotes the token unique identifier. Using the following relation, $activeDateTime$ and $expDateTime$ are computed.
\begin{IEEEeqnarray}{rCl}
	frdt &\leftarrow& finalReportDateTime(currentTime, \notag\\ &&freq_s, pd_s) \notag \\
	activeDateTime &\leftarrow&  frdt + tokActivDelay \notag \\
	expDateTime &\leftarrow& activeDateTime + validDays \label{eq:vas-protocol-details-token-generation}
\end{IEEEeqnarray}
The $finalReportDateTime$ function takes three parameters and computes the ending date-time of the final report $frdt$. The parameter $currentTime$ indicates the system's current time, the parameter $freq_s$ signifies the selected program's reporting frequency and the parameter $pd_s$ demonstrates the chosen program's duration. More specifically, the last report date-time of the selected program, $frdt$, is added to the token active delay time,  resulting in the active date-time of the generated token (i.e., $activeDateTime$). Subsequently, this resulting value is added to $validDays$,  yielding the token expiration date-time.
The parameter $value$ in $tok_s$ is computed in Relation \ref{eq:vas-protocol-details-value-spec} and the parameter $validDays$ is indicated in Relation \ref{eq:vas-protocol-details-validDays-spec}. The utility provider generates a unique ID $uid_s$ for the token $tok_s$ and signs it. 
\begin{IEEEeqnarray}{rCl}
	tok_s &=& \langle value, expDateTime, activeDateTime, uid_s\rangle \notag \\
	\sigma_{tok_s}^{UP} &\leftarrow& Sign(sk_{UP}^{\mathrm{sig}}, tok_s) 
	\label{eq:vas-protocol-details-sign-token}
\end{IEEEeqnarray}
The utility provider stores the token in its database. 

After the token generation, the utility provider signs the blinded credential $cr_{n-1, id_j}^{'}$ as follows: 
\begin{IEEEeqnarray}{rCl}
	\sigma_{cr_{n-1,id_j}^{'}}^{UP}
	&\leftarrow&
	BlindSign(sk_{UP}^{\mathrm{bs}},cr_{n-1,id_j}^{'})
	\label{eq:vas-protocol-details-sign-blinded-credential}
\end{IEEEeqnarray}
$BlindSign$ and $Unblind$ denote the RSA blind-signing and blind-factor removal operations defined in the preliminaries. The utility provider also signs $\sigma_{cr_{n-1, id_j}^{'}}^{UP} || tok_s || \sigma_{tok_s}^{UP}$ and then encrypts it. 
\begin{IEEEeqnarray}{rCl}
	\mathcal{M} &=& \sigma_{cr_{n-1, id_j}^{'}}^{UP} || tok_s || \sigma_{tok_s}^{UP} \notag \\
	\sigma_{\mathcal{M}}^{UP} &\leftarrow& Sign(sk_{UP}^{\mathrm{sig}}, \mathcal{M}) \notag \\ [2pt]
	\mathcal{X} &\leftarrow& Enc(pk_{id_j}, \mathcal{M} || \sigma_{\mathcal{M}}^{UP})
	\label{eq:vas-protocol-details-signcrypt-credential-token-tokenSignature}
\end{IEEEeqnarray}
The utility provider forwards $\mathcal{X}$ to the smart meter. 

\vspace{1mm}
\begin{normalsize}
	\noindent \textbf{Phase IV - Blind Factor Removal and Program-Specific MAC Key Generation:}
\end{normalsize}
The objective of this phase is to remove the blind factor from the last credential and obtain the utility provider's signature on that credential. Furthermore, a fresh program-specific MAC key is produced for consumption data integrity verification.

The smart meter receives $\mathcal{X}$ and decrypts it. It then verifies the signature $\sigma_{\mathcal{M}}^{UP}$.
\begin{IEEEeqnarray}{rCl}
	\mathcal{M}||\sigma_{\mathcal{M}}^{UP} &\leftarrow& Dec(sk_{id_j}, \mathcal{X}) \notag \\
	True/False &\leftarrow& Verify(pk_{UP}^{\mathrm{sig}}, \mathcal{M}, \sigma_{\mathcal{M}}^{UP})
	\label{eq:vas-protocol-details-decrypt-verify-token-credential}
\end{IEEEeqnarray}
Upon successful verification, the smart meter stores $tok_s$ and $\sigma_{tok_s}^{UP}$ securely. It then removes the blind factor $bf_{id_j}$ from signed blinded credential $\sigma_{cr_{n-1, id_j}^{'}}^{UP}$ to obtain $\sigma_{cr_{n-1, id_j}}^{UP}$.
\begin{IEEEeqnarray}{rCl}
	\sigma_{cr_{n-1,id_j}}^{UP}
	&\leftarrow&
	Unblind(\sigma_{cr_{n-1,id_j}^{'}}^{UP},bf_{id_j})
	\label{eq:vas-protocol-details-blind-factor-removal}
\end{IEEEeqnarray}
The smart meter ensures the integrity of the signature by applying the verification function as follows: 
\begin{IEEEeqnarray}{rCl}
	True/False &\leftarrow& Verify(pk_{UP}^{\mathrm{bs}}, cr_{n-1, id_j}, \sigma_{cr_{n-1, id_j}}^{UP})
 	\label{eq:vas-protocol-details-verify-the-correctness-of-signature}
\end{IEEEeqnarray}
If the verification fails, it means that the signature was tampered by illegitimate entities during transmission. According to this, the smart meter is able to detect the integrity violation and stop the process. But as we mentioned earlier, legitimate entities (e.g., aggregators and the utility provider) are semi-trusted, which means they follow the protocol and do not violate the integrity of message.

Afterward, the aggregator randomly selects one of the smart meters participating in program $pr_s$ (e.g., with $id_j$) and asks it to produce a fresh program-specific MAC key $k_{s}^{\mathrm{MAC}}$. This key is used by the utility provider, the aggregator, and the smart meters participating in $pr_s$ for consumption data integrity verification. Smart meters are trusted and tamper-proof, and we can rely on them to produce random MAC keys. The selected smart meter generates $k_{s}^{\mathrm{MAC}}$, signs the key, and then encrypts it with the public key of the aggregator.
\begin{IEEEeqnarray}{rCl}
	k_{s}^{\mathrm{MAC}} &\leftarrow& RanKeyGen(seed) \notag \\
	\sigma_{k_{s}^{\mathrm{MAC}}}^{id_j}
	&\leftarrow&
	Sign(sk_{id_j}, k_{s}^{\mathrm{MAC}}) \notag \\
	\mathsf{X}
	&\leftarrow&
	Enc(pk_{AGG},
	k_{s}^{\mathrm{MAC}} || \sigma_{k_{s}^{\mathrm{MAC}}}^{id_j})
	\label{eq:vas-protocol-details-gen-sharedKey-sign-encrypt}
\end{IEEEeqnarray}
The selected smart meter forwards $\mathsf{X}$ to the aggregator. 

\vspace{2mm}
\begin{normalsize}
	\noindent \textbf{Phase V - Program-Specific MAC Key Verification and Distribution:}
\end{normalsize}
In this phase, the aggregator verifies the received program-specific MAC key. Upon successful verification, it accepts the key and then distributes it to the utility provider and the remaining smart meters participating in program $pr_s$.

After receiving $\mathsf{X}$, the aggregator first decrypts and then verifies the program-specific MAC key $k_{s}^{\mathrm{MAC}}$.
\begin{IEEEeqnarray}{rCl}
	k_{s}^{\mathrm{MAC}} || \sigma_{k_{s}^{\mathrm{MAC}}}^{id_j}
	&\leftarrow&
	Dec(sk_{AGG}, \mathsf{X}) \notag \\ 
	True/False
	&\leftarrow&
	Verify(pk_{id_j}, k_{s}^{\mathrm{MAC}},
	\sigma_{k_{s}^{\mathrm{MAC}}}^{id_j}) \notag\\
	\label{eq:vas-protocol-details-agg-verify-shared-key}
\end{IEEEeqnarray}
Upon successful verification, it encrypts $k_{s}^{\mathrm{MAC}} || \sigma_{k_{s}^{\mathrm{MAC}}}^{id_j}$ using the public keys of the utility provider and the remaining smart meters participating in program $pr_s$, and then distributes the program-specific MAC key among them. Each receiving entity can verify the key to ensure that it originates from a trusted meter and has not been tampered with by semi-trusted or untrusted entities. Now, the utility provider, the aggregator, and the participating smart meters have the program-specific MAC key $k_{s}^{\mathrm{MAC}}$ for data integrity checking during program $pr_s$.

\vspace{1mm}
\begin{normalsize}
	\noindent \textbf{Phase VI - Anonymous Metering Data Reporting:}
\end{normalsize}
During this phase, the meter applies a temporal-based aggregation scheme to produce coarser-grained consumption data. It then applies a local differential privacy mechanism based on the predefined privacy budget and the selected noise scale to generate noisy coarse-grained consumption data. Afterward, the meter computes the MAC of the noisy coarse-grained consumption data for later integrity verification. Finally, it reports encrypted readings to the utility provider via an anonymous overlay network.

At this point, the participating smart meter has obtained $tok_i$, $\sigma_{tok_i}^{UP}$, $\sigma_{cr_{n-1, id_j}}^{UP}$, and the program-specific MAC key $k_{s}^{\mathrm{MAC}}$. The smart meter aggregates measured consumption values to generate a coarse-grained consumption value (i.e., temporal-based aggregation). If the selected program has $nsc_s>0$, the meter applies the local differential privacy mechanism to the clipped coarse-grained consumption value using the predefined privacy parameters  $(\epsilon,\delta)$ and the selected noise-scale parameter $nsc_s$. If $nsc_s=0$, the meter forwards the clipped coarse-grained consumption value without adding
Gaussian noise; therefore, no local differential privacy guarantee is provided.

Let $P_{\max}$ denote the maximum power permitted by the household's electrical connection and $T_s$ denote the reporting-window duration of program $pr_s$. The upper bound is predefined as $C_s=P_{\max}T_s$, where $C_s$ represents the maximum possible coarse-grained consumption sum during that window. The smart
meter first clips the coarse-grained consumption value to the bounded domain $[0,C_s]$ as follows:
\begin{IEEEeqnarray}{rCl}
	c_{t_i}^{B}
	&\leftarrow&
	\min\{\max\{c_{t_i},0\},C_s\}
	\label{eq:vas-clipped-coarse-value}
\end{IEEEeqnarray}

Because the released function is the identity function over the bounded
domain $[0,C_s]$, its $L_2$-sensitivity is defined as follows:
\begin{IEEEeqnarray}{rCl}
	\Delta_{2,s}
	&=&
	\max_{c,c'\in[0,C_s]} |c-c'|
	= C_s
	\label{eq:vas-coarse-sensitivity}
\end{IEEEeqnarray}
The two values $c$ and $c'$ are two alternative possible values for the same report. They are not necessarily two consecutive readings in time.

If the noise scale is set to one (i.e., the default noise level), the
protocol applies the Gaussian mechanism to the clipped coarse-grained
consumption value. For $0<\epsilon<1$ and $0<\delta<1$, the baseline
noise addition is defined as follows:
\begin{IEEEeqnarray}{rCl}
	\sigma_s
	&=&
	\dfrac{\Delta_{2,s}\sqrt{2\ln(1.25/\delta)}}{\epsilon}
	\notag\\
	\xi_{t_i}
	&\sim&
	\mathscr{N}(0,\sigma_s^2)
	\notag\\
	\tilde{c}_{t_i}
	&\leftarrow&
	c_{t_i}^{B}+\xi_{t_i}
	\label{eq:vas-protocol-details-baseline-noise-definition}
\end{IEEEeqnarray}
Where:
\begin{enumerate}
	\item $c_{t_i}$ indicates the coarse-grained consumption sum at interval $t_i$.
	\item $c_{t_i}^{B}$ represents the coarse-grained consumption sum after clipping it to the domain $[0,C_s]$.
	\item $\Delta_{2,s}$ denotes the $L_2$-sensitivity of the coarse-grained consumption value for program $pr_s$.
	\item $\epsilon$ is the privacy budget, controlling the trade-off between privacy and data utility.
	\item $\delta$ bounds the probability of exceptional privacy loss.
	\item $\sigma_s$ represents the baseline standard deviation of the Gaussian noise.
	\item $\xi_{t_i}$ is a random variable sampled independently from a normal distribution with mean zero and variance $\sigma_s^2$.
	\item $\tilde{c}_{t_i}$ is the perturbed coarse-grained consumption value.
\end{enumerate}

The smart meter applies the chosen noise scale parameter $nsc_s$ to  adjust the baseline noise scale as follows for $nsc_s>0$:
\begin{IEEEeqnarray}{rCl}
	\hat{\sigma}_s
	&=&
	nsc_s\cdot\sigma_s
	\notag\\
	\hat{\xi}_{t_i}
	&\sim&
	\mathscr{N}(0,\hat{\sigma}_s^2)
	\notag\\
	\tilde{c}_{t_i}
	&\leftarrow&
	c_{t_i}^{B}+\hat{\xi}_{t_i}
	\label{eq:vas-protocol-details-applying-scale-parameter}
\end{IEEEeqnarray}

For $nsc_s>0$, the effective per-report privacy budget of program $pr_s$ is defined as follows:
\begin{IEEEeqnarray}{rCl}
	\epsilon_s
	&=&
	\dfrac{\epsilon}{nsc_s}
	\label{eq:vas-effective-privacy-budget}
\end{IEEEeqnarray}
Consequently, each perturbed report satisfies $(\epsilon_s,\delta)$-local differential privacy, where $\epsilon_s=\epsilon/nsc_s$, provided that $0<\epsilon_s<1$. Choosing programs with $nsc_s>1$ results in $\epsilon_s<\epsilon$, thereby providing stronger privacy. Conversely, choosing programs with $0<nsc_s<1$ results in $\epsilon_s>\epsilon$, thereby improving accuracy but providing a weaker privacy guarantee. When the perturbation mechanism is applied repeatedly to the same user's reports, the privacy losses accumulate according to the basic sequential composition property of differential privacy \cite{dwork2014algorithmic}. Therefore, for a program that produces $q_s=freq_s\cdot pd_s$ reports, each satisfying $(\epsilon_s,\delta)$-LDP, the complete reporting session satisfies $(q_s\epsilon_s,q_s\delta)$-LDP, where the system parameters are selected such that $q_s\delta<1$. If $nsc_s=0$, the Gaussian noise sampling step is omitted, $\tilde{c}_{t_i}\leftarrow c_{t_i}^{B}$, and the program provides no local differential privacy guarantee.

The smart meter applies a data minimization technique based on the data collection purpose (denoted by $prp_i$ in the program $pr_i$) ensuring that only the necessary fields required for the selected service or purpose are forwarded while irrelevant data is omitted. According to Tudor et al. \cite{tudor2013analysis}, metering data used for various purposes such as billing and operations require different fields. Therefore, the adoption of an appropriate data minimization technique is recommended.

Let $e\in\{0,\ldots,pd_s-1\}$ denote the daily reporting epoch and $r\in\{0,\ldots,freq_s-1\}$ denote the report sequence number within that epoch. For the first report, $(e,r)=(0,0)$.

For each program session, the smart meter independently generates a fresh pseudonym $uuid_{\omega}$. Afterward, the smart meter replaces its identity with the pseudonym $uuid_{\omega}$ and computes the MAC of  $pid_s||e||r||\tilde{c}_{t_{0,uuid_{\omega}}}||uuid_{\omega}$ using the program-specific MAC key $k_{s}^{\mathrm{MAC}}$ (denoted by $tag_{\upsilon_{0,\omega}}$) as follows:
\begin{IEEEeqnarray}{rCl}
	tag_{\upsilon_{0,\omega}}
	&\leftarrow&
	\text{MAC}(k_{s}^{\mathrm{MAC}},
	pid_s || e || r \notag\\
	&& {} || \tilde{c}_{t_{0,uuid_{\omega}}} || uuid_{\omega})
	\label{eq:vas-protocol-details-applying-MAC-on-consumption-uuid}
\end{IEEEeqnarray}
All protocol fields, including consumption values, are encoded using fixed-length binary representations. The computed tag $tag_{\upsilon_{0,\omega}}$ enables the utility provider to check the integrity of the received consumption values, reporting epoch, and sequence number during the program execution. Subsequently, the smart meter forwards
$pid_s||e||r||cr_{n-1,uuid_{\omega}}||
\sigma_{cr_{n-1,uuid_{\omega}}}^{UP}||
\tilde{c}_{t_{0,uuid_{\omega}}}||uuid_{\omega}||
tag_{\upsilon_{0,\omega}}$
in an encrypted format (using the utility provider's encryption public key)
to the utility provider via an anonymous overlay network.
\begin{IEEEeqnarray}{rCl}
	\mathscr{M}
	&=& pid_s || e || r || cr_{n-1,uuid_{\omega}}
	|| \sigma_{cr_{n-1,uuid_{\omega}}}^{UP} \notag\\
	&& {} || \tilde{c}_{t_{0,uuid_{\omega}}}
	|| uuid_{\omega} || tag_{\upsilon_{0,\omega}} \notag\\
	\mathscr{X}
	&\leftarrow&
	Enc(pk_{UP}^{\mathrm{enc}},\mathscr{M})
	\label{eq:vas-protocol-details-anonymous-reporting}
\end{IEEEeqnarray}

\begin{normalsize}
	\noindent \textbf{Phase VII - Credential and Meter Data Verification and Secure Archiving:}
\end{normalsize}
This phases ensures the verification of forwarded anonymous credentials and consumption data at each reporting interval. Upon successful verification, the utility provider updates the last recorded credential and securely stores the new consumption data.

The utility provider, first decrypts $\mathscr{X}$ and verifies the credential signature $\sigma_{cr_{n-1, uuid_{\omega}}}^{UP}$ as follows: 
\begin{IEEEeqnarray}{rCl}
	\mathscr{M} &\leftarrow& Dec(sk_{UP}^{\mathrm{enc}}, \mathscr{X}) \notag \\
	True/False &\leftarrow& Verify(pk_{UP}^{\mathrm{bs}}, cr_{n-1, uuid_{\omega}}, \sigma_{cr_{n-1, uuid_{\omega}}}^{UP})\notag \\
	\label{eq:vas-protocol-details-verify-credential-signature}
\end{IEEEeqnarray}
For the first report, where $(e,r)=(0,0)$, upon successful credential verification, the utility provider checks the integrity of $pid_s||e||r||\tilde{c}_{t_{0,uuid_{\omega}}}||uuid_{\omega}$ using the program-specific MAC key $k_{s}^{\mathrm{MAC}}$.
\begin{IEEEeqnarray}{rCl}
	tag_{\upsilon_{0,\omega}}^{'}
	&\leftarrow&
	\text{MAC}(k_{s}^{\mathrm{MAC}},
	pid_s || e || r \notag\\
	&& {} || \tilde{c}_{t_{0,uuid_{\omega}}} || uuid_{\omega}) \notag \\
	True/False
	&\leftarrow&
	\text{EqCheck}(tag_{\upsilon_{0,\omega}},
	tag_{\upsilon_{0,\omega}}^{'})
	\label{eq:vas-protocol-details-credential-tag-verification}
\end{IEEEeqnarray}
Upon successful verification, the utility provider stores the current credential $cr_{n-1,uuid_{\omega}}$ and the consumption value $\tilde{c}_{t_{0,uuid_{\omega}}}$ under $pid_s$ and the corresponding reporting position $(e,r)$.
After receiving a positive acknowledgment for the first report, the smart meter advances to the next credential. For the second report, $(e,r)=(0,1)$, and the meter forwards $pid_s||e||r||cr_{n-2,uuid_{\omega}}|| 
\tilde{c}_{t_{1,uuid_{\omega}}}||uuid_{\omega}||tag_{\upsilon_{1,\omega}}$ to the utility provider in encrypted form through the anonymous overlay network, where $tag_{\upsilon_{1,\omega}}$ is computed over $pid_s||e||r||\tilde{c}_{t_{1,uuid_{\omega}}}||uuid_{\omega}$ using $k_s^{\mathrm{MAC}}$. The utility provider verifies the credential as follows:
\begin{IEEEeqnarray}{rCl}
	cr_{n-1,uuid_{\omega}}^{'}
	&\leftarrow&
	H(cr_{n-2,uuid_{\omega}}) \notag \\
	True/False
	&\leftarrow&
	\text{EqCheck}(cr_{n-1,uuid_{\omega}},
	cr_{n-1,uuid_{\omega}}^{'})
	\label{eq:vas-protocol-details-credential-verfication-using-haschain-function}
\end{IEEEeqnarray}
If $cr_{n-1,uuid_{\omega}}^{'}$ matches $cr_{n-1,uuid_{\omega}}$, the utility provider stores $cr_{n-2,uuid_{\omega}}$ as the latest accepted credential. Similar to Relation \ref{eq:vas-protocol-details-credential-tag-verification}, it
verifies the integrity of $pid_s||e||r||\tilde{c}_{t_{1,uuid_{\omega}}}||uuid_{\omega}$ using the program-specific MAC key $k_s^{\mathrm{MAC}}$. The same credential and consumption-data validation is performed for subsequent reports. Within each daily epoch, $r$ is incremented for each successfully acknowledged report; after $r=freq_s-1$, the next report starts the following epoch with $(e,r)=(e+1,0)$.

\vspace{1mm}
\begin{normalsize}
	\noindent \textbf{Anonymous Acknowledgment and Loss-Resilient Reporting:}
\end{normalsize}
For each successfully accepted report, the utility provider and the smart meter independently derive the acknowledgment identifier
\begin{IEEEeqnarray}{rCl}
	ack_{e,r,\omega}
	&\leftarrow&
	H(\text{"ACK"} || pid_s || e || r || uuid_{\omega} \notag\\
	&& {} || cr_{n-1-(e\cdot freq_s+r),uuid_{\omega}})
	\label{eq:vas-anonymous-ack}
\end{IEEEeqnarray}
The utility provider inserts $ack_{e,r,\omega}$ into the Bloom filter
$BF_{e,r}$ corresponding to reporting slot $(e,r)$, signs
$pid_s||e||r||BF_{e,r}$, and broadcasts the signed filter through the
aggregator to the participating meters. The filter is also broadcast when
no report is accepted during the corresponding slot, in which case it
remains empty. Each meter verifies the utility provider's signature and
checks its locally derived acknowledgment identifier in $BF_{e,r}$. The
meter advances to the next report sequence and credential only after a
positive acknowledgment; otherwise, after a predefined timeout
$\tau_{\mathrm{ack}}$, it retransmits the same report with the same $e$,
$r$, and credential. Duplicate retransmissions are discarded by the
utility provider without advancing its credential state.

For a Bloom filter with $m$ bits, $n$ inserted acknowledgment identifiers,
and $k$ hash functions, the false-positive probability can be approximated
as $P_{\mathrm{FP}}\approx(1-e^{-kn/m})^k$
\cite{grandi2018analysis}. In the largest considered setting of 1,000
participating meters, 16 reports per day, and 21 days, the program contains
$21\times16=336$ reporting slots and 336,000 reports overall. Since each
$BF_{e,r}$ contains at most 1,000 acknowledgment identifiers, allocating
24 bits per expected entry gives $m=24000$ bits (approximately 3 KB).
Using the near-optimal $k=17$ hash functions results in
$P_{\mathrm{FP}}\approx9.84\times10^{-6}$. Thus, the mechanism provides
resilience against transient report loss or message dropping without
requiring a direct acknowledgment path to the originating meter.

\begin{center}
	\begin{figure*}[htp]
		\centering
		\includegraphics[width=5.5in]{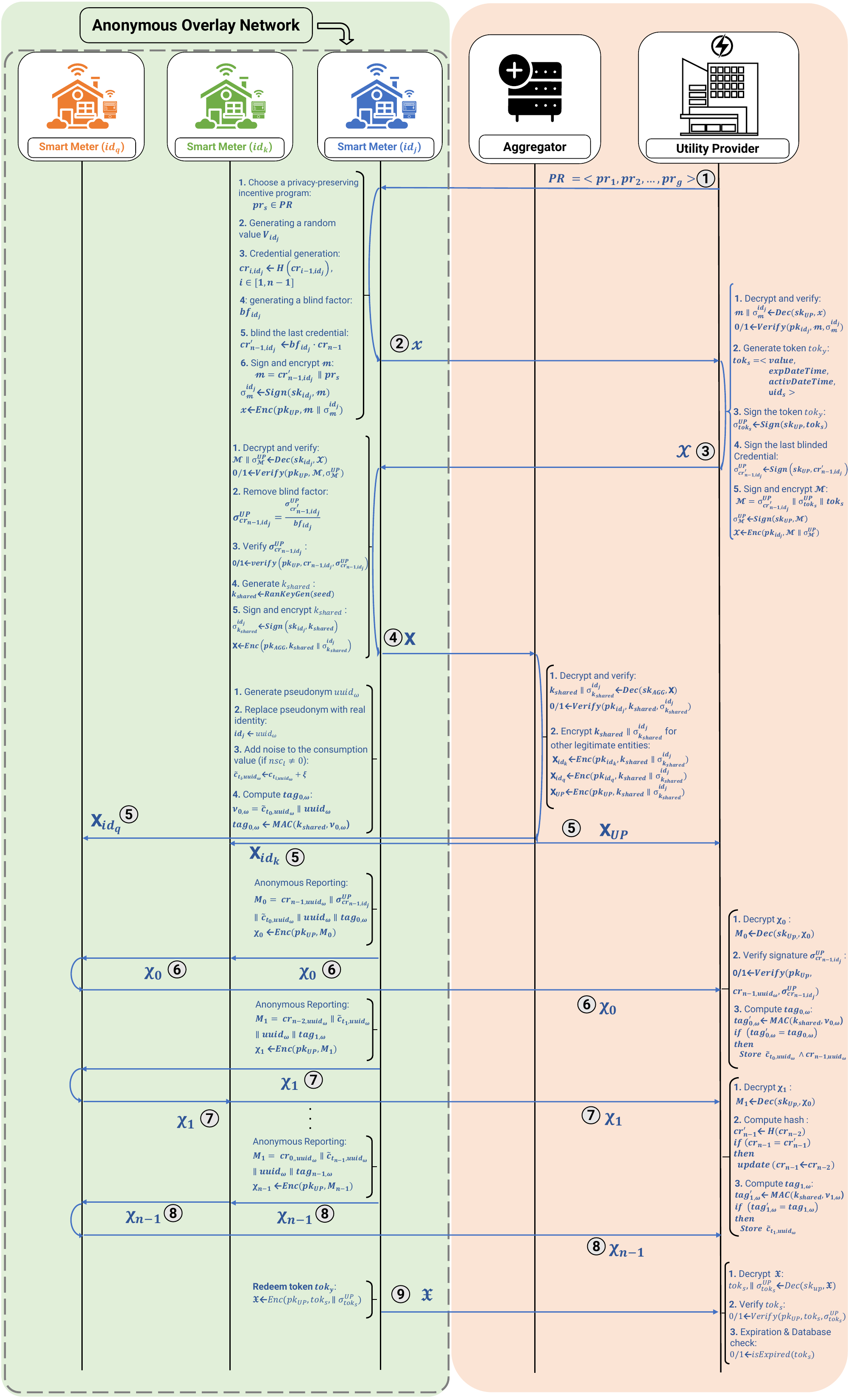}
		\caption{The details of the protocol flow, which consists of various phases, are illustrated. The smart with $id_j$ participates in the privacy incentive program $pr_s$ and executes the protocol phases. Additionally, it serves as the designated smart meter responsible for generating program-specific MAC key used for data integrity verification. Smart meters with $id_{k}$ and $id_q$ are also participate in the program $pr_s$ and are considered as neighboring meters that act as relaying nodes for the smart meter with $id_j$ to conceal data source.}
		\label{vas-protocol-details}
	\end{figure*}
\end{center}

\begin{normalsize}
	\noindent \textbf{Phase VIII - Token Redemption:}
\end{normalsize}
In the end, the smart meter activates the token for the customer and enables him/her to spend the token in the energy infrastructure (e.g., billing). When a user redeems the token, the utility provider first verifies the signature as follows: 
\begin{IEEEeqnarray}{rCl}
	True/False &\leftarrow&  Verify(pk_{UP}^{\mathrm{sig}}, tok_{i}, \sigma_{tok_i}^{UP}) \notag \\
	True/False &\leftarrow& \text{isExpired}(tok_i)
	\label{eq:vas-protocol-details-credential-verify-token-check-expiray-date}
\end{IEEEeqnarray}
Upon successful verification, the utility provider checks its database to ensure that the token has not been used before, preventing double-spending. if the token has not been used, the utility provider grants permission to redeem it. Afterward, it updates the status field in the database to "spent" and prevents any further use of the token. \\
The protocol preserves users' privacy while collecting customers' consumption values for useful purposes such as data analysis, machine learning, and other data-driven applications. The protocol's flow is depicted in Fig. \ref{vas-protocol-details}.

\section{Evaluation}
We evaluate our proposed protocol from two perspectives. First, we examine  performance metrics, including computational, memory, and communication overhead. Specifically, we provide an analytical evaluation for all the three metrics and supplement it with an experimental evaluation of the computational overhead to demonstrate the scheme's efficiency. Second, we conduct a privacy and security analysis, demonstrating that the scheme preserves customer privacy against both legitimate semi-trusted entities and untrusted adversaries.
\subsection{Performance Evaluation}
The protocol ran on two separate machines. The first machine was an Orange Pi One PC  with a 1.2GHz 32-bit ARM processor (Cortex A7) and 1GB of RAM, used to simulate a smart meter. We utilized Quick Emulator (QEMU) to simulate the Orange Pi One PC board. The second machine featured a 1.8GHz core i7 intel processor with 8 cores and 32GB of RAM, used to simulate the aggregator and the utility provider. To demonstrate the feasibility of the protocol under limited memory conditions, we restricted the RAM of the second machine to 1GB. \\
The first machine ran an Armbian operating system (Ubuntu-based), which is optimized for IoT applications, whereas the second machine operates on a standard Debian12 operating system. 

The entire protocol has been implemented in Python using optimized cryptographic libraries for both cryptographic primitives and the data perturbation algorithm utilized in the proposed protocol. 

We employed a dataset \cite{cenky2023dataset} sourced from Slovakia's AMI infrastructure, which comprises energy consumption records from 1,000 anonymized residential smart meters. This dataset provides both active and reactive energy measurements at 15-minute intervals throughout an entire year. It has been secured using conventional privacy techniques like data minimization and anonymization, and notably, it is free of null values and data sparsity issues.

In the following, the evaluation results of the key performance metrics are reported.


\subsubsection{Computational Overhead} Smart meters have constrained resources, and many use 32-bit ARM-based processors with low clock rates. To have an insight into the scheme's computational overhead, we estimate this metric for our scheme. 

For the analytical evaluation of the protocol's computational overhead, we analyze the computations performed by each legitimate entity separately.  We provide a notation table (see Table \ref{table:vas-evaluation-computational-overhead-notation}) for the symbols that have been used for the analytical evaluation.

\begin{table}[htp]
	\centering
	\caption{Notation Table for Computational Overhead}
	\renewcommand{\arraystretch}{1.5} 
	\setlength{\tabcolsep}{6pt} 
	\resizebox{1.01\columnwidth}{!}{ 
		\begin{tabular}{|p{3.5cm}|p{9cm}|}
			\hline
			\textbf{Symbol} & \textbf{Description} \\
			\hline
			$\mathbf{t}_{PR}$ & Time of generating a list of privacy-preserving incentive programs \\
			\hline
			$\mathbf{t}_{RandGen}$ & Time of generating random values such as blind factor, uuid, or noise\\
			\hline
			$\mathbf{t}_{arthm}$ & Time of performing arithmetic operations, including adding noise to coarse-grained consumption readings, blinding the last credential, removing the blind factor from the last credential, or performing equality check operations\\
			\hline
			$\mathbf{t}_{asym-keyGen}$ & Time of generating cryptographic keys for an asymmetric cryptosystem\\
			\hline
			$\mathbf{t}_{sym-keyGen}$ & Time of generating a program-specific MAC key\\
			\hline
			$\mathbf{t}_{asym-op}$ & Time of performing various asymmetric cryptographic operations, including encryption, decryption, signing, and verification \\
			\hline
			$\mathbf{t}_{MAC}$ & Time of computing MAC of a data\\
			\hline
			$\mathbf{t}_{tokGen}$ & Time of generating a token\\
			\hline
			$\mathbf{t}_{DB}$ & Time of performing database operations such as select, insert, or update\\
			\hline
			$\mathbf{t}_{Hash}$ & Time of computing a hash value \\
			\hline
		\end{tabular}
	}
	\label{table:vas-evaluation-computational-overhead-notation}
\end{table}

The computational overhead of the smart meter in the worst-case scenario for the complete execution of one privacy-preserving incentive program is
given as follows, where $n$ denotes the total number of coarse-grained anonymous reports generated during that program execution.

\begin{IEEEeqnarray}{rCl}
	\mathbf{t}_{sm} &=& 
	(n+8)\mathbf{t}_{asym-op} + (n+3)\mathbf{t}_{RandGen}
	\notag \\[1pt] 
	& & +  (n-1) \mathbf{t}_{Hash} + (n+2)t_{arthm} + n\mathbf{t}_{MAC}\notag \\[1pt]
	& & +  \mathbf{t}_{keyGen-asym}  + \mathbf{t}_{keyGen-sym} 	\label{eq:vas-evaluation-computational-overhead-SM}
\end{IEEEeqnarray}
For instance, $(n+8)\mathbf{t}_{asym-op}$ indicates that the smart meter performs $n+8$ asymmetric operations, such as encryption, decryption, signing, and verification during the protocol execution.

The aggregator's computational overhead is less than the other two entities. The analytical computational overhead of the aggregator is as follows where $\mathscr{N}$ is the number of participating meters in the program: 

\begin{IEEEeqnarray}{rCl}
	\mathbf{t}_{agg} &=& (\mathscr{N} + 2) \mathbf{t}_{asym-op} + \mathbf{t}_{KeyGen-asym} 
	\label{eq:vas-evaluation-computational-overhead-AGG}
\end{IEEEeqnarray}
The aggregator's cryptographic operations mainly consist of selecting a participating smart meter to generate the program-specific MAC key
$k_s^{\mathrm{MAC}}$, verifying the received key, and distributing it to the utility provider and the remaining smart meters participating in
program $pr_s$. Next, we examine the utility provider's computational overhead.

\begin{IEEEeqnarray}{rCl}
	\mathbf{t}_{up} &=& 
	\mathbf{t}_{PR} + (n+12)\mathbf{t}_{asym-op} + \mathbf{t}_{tokGen} + n\mathbf{t}_{MAC}   \notag \\[2pt]
	& & + (2n-1)\mathbf{t}_{arthm} + (n-1)\mathbf{t}_{Hash} + (3n-1) \mathbf{t}_{DB}   \notag \\[2pt]
	& &  
	+ \mathbf{t}_{keyGen-asym}
	\label{eq:vas-evaluation-computational-overhead-UP}
\end{IEEEeqnarray}

Finally, the computational overhead of the protocol is as follows:
\begin{IEEEeqnarray}{rCl}
	\mathbf{t}_{vas} &=& \mathbf{t}_{sm} + \mathbf{t}_{agg} + \mathbf{t}_{up}
	\label{eq:vas-evaluation-computational-overhead}
\end{IEEEeqnarray}

We conduct an experiment to measure the protocol's execution time. Table \ref{table:vas-evaluation-computational-overhead-experimental} summarizes
the protocol's runtime. As mentioned earlier, the aggregator functions mainly as an intermediary node to distribute the program-specific MAC key
among the corresponding participants and to forward consumption data. Based on this, the core of the protocol is running on the smart meter and
utility provider's side.

\begin{center}
	\begin{figure}[htp]
		\centering
		\includegraphics[width=3.6in]{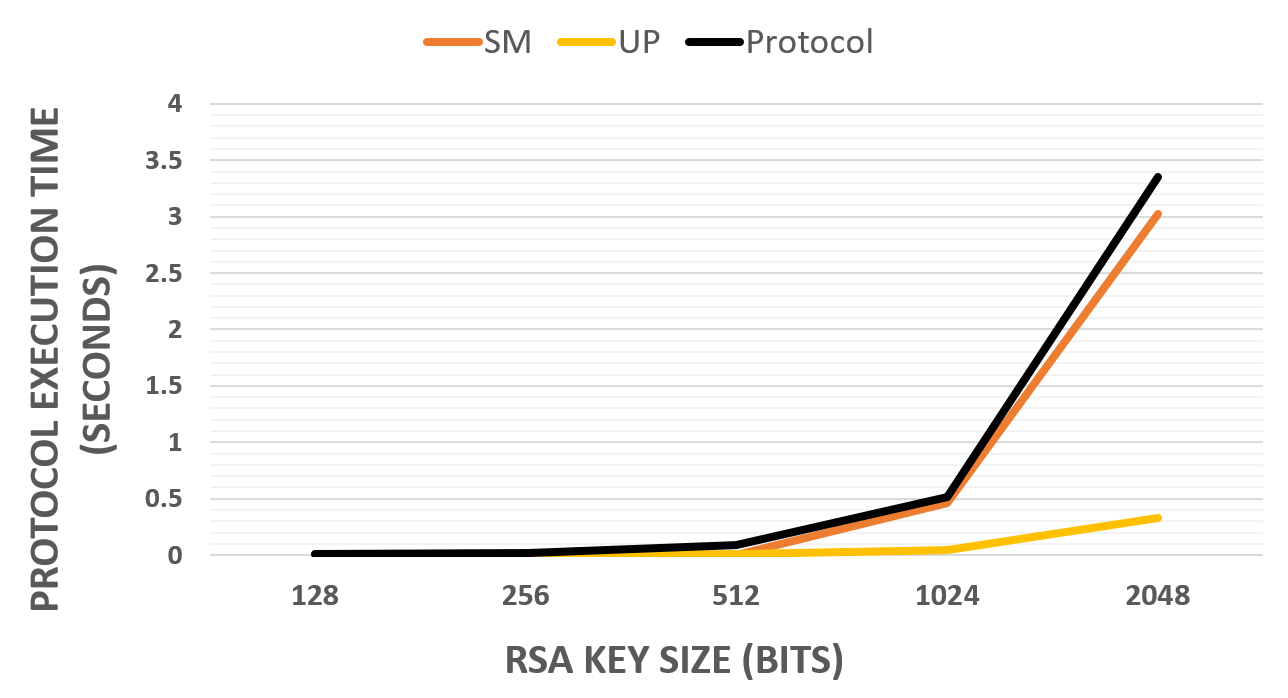}
		\caption{The scheme's execution time based on different RSA key size}
		\label{vas-protocol-execution-time}
	\end{figure}
\end{center}

\begin{table}[htp]
	\centering
	\caption{Protocol Execution Time for Different Entities in the Value-Added Service (Frequency: 4 – Duration: 7)}
	\normalsize
	\renewcommand{\arraystretch}{1.5}
	\setlength{\tabcolsep}{6pt}
	\resizebox{1.00\columnwidth}{!}{
		\begin{tabular}{|c|c|c|c|}
			\hline
			\multirow{2}{*}{\textbf{RSA Key Length}} & \multicolumn{3}{c|}{\textbf{Execution Time (in seconds) for Different Entities}} \\
			\cline{2-4}
			& \textbf{Smart Meter ($t_{sm}$)} & \textbf{Utility Provider ($t_{up}$)} & \textbf{Total Execution Time ($t_{vas}$)} \\
			\hline
			\textbf{128-bit} &  0.00875 & 0.00117 & 0.00992 \\
			\hline
			\textbf{256-bit} & 0.01950 & 0.00216  & 0.02166\\
			\hline
			\textbf{512-bit} & 0.07977 & 0.00941 & 0.08918 \\
			\hline
			\textbf{1024-bit} & 0.46193 & 0.04914 & 0.51107 \\
			\hline
			\textbf{2048-bit} & 3.02650 &  0.32894  & 3.35544\\
			\hline
		\end{tabular}
	}
	\label{table:vas-evaluation-computational-overhead-experimental}
\end{table}
As illustrated in Fig. \ref{vas-protocol-execution-time}, the execution time increases with the RSA key size. For the commonly used 2048-bit RSA
key length, the complete execution of the protocol for a 7-day program with four reports per day takes approximately $3.36$ seconds, including
$3.03$ seconds at the smart meter and $0.33$ seconds at the utility provider. In this setting, the program contains $n=4\times7=28$ coarse-grained reports. Therefore, when the total execution cost is amortized over the complete reporting session, the average computational overhead is approximately $0.108$ seconds per report at the smart meter, $0.012$ seconds per report at the utility provider, and $0.120$ seconds per report for the overall protocol. These values are small relative to the reporting interval; for four reports per day, consecutive reports are separated by approximately six hours. Since the proposed value-added service does not require real-time reporting, this computational delay does not affect the practical operation of the protocol.

\subsubsection{Memory Overhead}
Memory overhead is another essential metric. In our protocol, each legitimate entity uses 1GB of RAM to perform various operations such as signing, verification, encryption, decryption, noise generation, and noise addition. Based on our experiment, in the worst-case scenario, the complete protocol execution with a 2048-bit RSA key size, a 7-day program duration, and four reports per day consumes approximately 4.5 MB of memory on average, making it suitable for memory-constrained environments. 

\subsubsection{Communication Overhead} Finally, let's examine the scheme's communication overhead. Similarly, we provide a notation table (see Table \ref{table:vas-evaluation-communication-overhead-notation}) for the communication overhead. 
\begin{table}[htp]
	\centering
	\caption{Notation Table for Communication Overhead}
	\renewcommand{\arraystretch}{1.5} 
	\setlength{\tabcolsep}{6pt} 
	\resizebox{1.00\columnwidth}{!}{ 
		\begin{tabular}{|p{3.5cm}|p{9cm}|}
			\hline
			\textbf{Symbol} & \textbf{Description} \\
			\hline
			$\mathbf{P}_{PR}$ & The data size for the list of privacy-preserving incentive programs \\
			\hline
			$\mathbf{P}_{\mathcal{x}}$ & The data size for the encrypted message $\mathcal{x}$ \\
			\hline
			$\mathbf{P}_{\mathcal{X}}$ & The data size for the encrypted message $\mathcal{X}$  \\
			\hline
			$\mathbf{P}_{\mathsf{X}}$ & The data size for the encrypted message $\mathsf{X}$ \\
			\hline
			$\mathbf{P}_{\mathscr{X}_0}$ & The data size for the first encrypted message $\mathscr{X}_0$ \\
			\hline
			$\mathbf{P}_{\mathscr{X}_i}$ & The data size for the subsequent encrypted message $\mathscr{X}_i$ \\
			\hline
			$\mathbf{P}_{\mathfrak{X}}$ & The data size for the encrypted message  $\mathfrak{X}$ \\
			\hline
		\end{tabular}
	}
	\label{table:vas-evaluation-communication-overhead-notation}
\end{table}

We evaluate the communication overhead in both Neighborhood Area Network (NAN) and Wide Area Network (WAN) based on the adopted link layer technologies. In the proposed network model, smart meters send metering data using 6LoWPAN and IEEE 802.15.4g (Wi-SUN) to the aggregator. IEEE 802.15.4g is standardized for NAN to connect smart meters to intermediary aggregators \cite{chang2012ieee}. Afterward, the aggregator relays the received metering data to the utility provider via 4G-LTE infrastructure. Specifically, the metering data first is transmitted to the 4G-LTE Radio Access Network (RAN) or Evolved Node B (eNB), then routed to Packet Data Network Gateway (PGW), which ultimately forwards it to the utility provider. Table \ref{table:network-packet-bandwidth} presents communication overhead for each link in the network. Since each link employs a distinct network stack, the associated overhead varies. As shown, the minimum required bandwidth is 320 kbps for Wi-SUN and 240 kbps for both 4G-LTE and Ethernet—rates that are easily supported by these  wireless and wired communication technologies. IEEE 802.15.4g provides bandwidth of at least 40 kbps to at most 1000 kbps \cite{chang2012ieee}. We recommend using 802.11ah as NAN technology which provides a bandwidth of 78 Mbps and range of 1000 m \cite{ahmed2016comparison}. 802.11ah can provide better bandwidth but currently is not as widely adopted by utility providers (or grid infrastructure) as 802.15.4g does, due to the cost of new technology deployment.
\textbf{\begin{table*}[htp]
		\centering
		\caption{Packet sizes and per-meter transmission times (20 SMs, 12.5 kbps BW per meter in NAN, 50 kbps BW per meter in WAN)}
		\renewcommand{\arraystretch}{1.5}
		\setlength{\tabcolsep}{6pt}
		\resizebox{\textwidth}{!}{%
			\begin{tabular}{|>{\centering\arraybackslash}p{4.5cm}|c|c|c|c|c|c|c|c|}
				\hline
				\multirow{3}{*}{\textbf{Payload}} 
				& \multicolumn{2}{c|}{\textbf{NAN}} 
				& \multicolumn{6}{c|}{\textbf{WAN}} \\ \cline{2-9}
				& \multicolumn{2}{c|}{%
					\begin{tabular}[c]{@{}c@{}}\textbf{SM $\leftrightarrow$ AGG}\\
						Wireless – IEEE 802.15.4g (Wi-SUN)\end{tabular}}
				& \multicolumn{2}{c|}{%
					\begin{tabular}[c]{@{}c@{}}\textbf{AGG $\leftrightarrow$ eNB}\\
						Wireless – 4G-LTE (PDCP-LTE)\end{tabular}}
				& \multicolumn{2}{c|}{%
					\begin{tabular}[c]{@{}c@{}}\textbf{eNB $\leftrightarrow$ PGW}\\
						Wired – IEEE 802.3 (Ethernet)\end{tabular}}
				& \multicolumn{2}{c|}{%
					\begin{tabular}[c]{@{}c@{}}\textbf{PGW $\leftrightarrow$ UP}\\
						Wired – IEEE 802.3 (Ethernet)\end{tabular}} \\ \cline{2-9}
				& \textbf{Packet Size} & \textbf{Trans. Time} 
				& \textbf{Packet Size} & \textbf{Trans. Time} 
				& \textbf{Packet Size} & \textbf{Trans. Time} 
				& \textbf{Packet Size} & \textbf{Trans. Time} \\ \hline
				$P_{PR}=5{,}480\,\text{B}$ (10 programs) 
				& $7{,}099\,\text{B}$ & $4.54336\,\text{s}$
				& $5{,}534\,\text{B}$ & $0.88544\,\text{s}$
				& $5{,}992\,\text{B}$ & $0.95872\,\text{s}$
				& $5{,}768\,\text{B}$ & $0.92288\,\text{s}$ \\ \hline
				$P_{\mathcal{x}}=1{,}024\,\text{B}$ 
				& $1{,}323\,\text{B}$ & $0.84672\,\text{s}$
				& $1{,}078\,\text{B}$ & $0.17248\,\text{s}$
				& $1{,}146\,\text{B}$ & $0.18336\,\text{s}$
				& $1{,}090\,\text{B}$ & $0.17440\,\text{s}$ \\ \hline
				$P_{\mathcal{X}}=768\,\text{B}$ 
				& $1{,}007\,\text{B}$ & $0.64448\,\text{s}$
				& $822\,\text{B}$   & $0.13152\,\text{s}$
				& $890\,\text{B}$   & $0.14240\,\text{s}$
				& $834\,\text{B}$   & $0.13344\,\text{s}$ \\ \hline
				$P_{\mathsf{X}}=512\,\text{B}$ 
				& $661\,\text{B}$ & $0.42304\,\text{s}$
				& $566\,\text{B}$ & $0.09056\,\text{s}$
				& $634\,\text{B}$ & $0.10144\,\text{s}$
				& $578\,\text{B}$ & $0.09248\,\text{s}$ \\ \hline
				$P_{\mathscr{X}_0}=512\,\text{B}$ 
				& $661\,\text{B}$ & $0.42304\,\text{s}$
				& $566\,\text{B}$ & $0.09056\,\text{s}$
				& $634\,\text{B}$ & $0.10144\,\text{s}$
				& $578\,\text{B}$ & $0.09248\,\text{s}$ \\ \hline
				$P_{\mathscr{X}_i}=256\,\text{B}$ 
				& $345\,\text{B}$ & $0.22080\,\text{s}$
				& $345\,\text{B}$ & $0.05520\,\text{s}$
				& $345\,\text{B}$ & $0.05520\,\text{s}$
				& $345\,\text{B}$ & $0.05520\,\text{s}$ \\ \hline
				$P_{\mathfrak{X}}=512\,\text{B}$ 
				& $661\,\text{B}$ & $0.42304\,\text{s}$
				& $566\,\text{B}$ & $0.09056\,\text{s}$
				& $634\,\text{B}$ & $0.10144\,\text{s}$
				& $578\,\text{B}$ & $0.09248\,\text{s}$ \\ \hline
				\textbf{Available BW} 
				& \multicolumn{2}{c|}{250 kbps} 
				& \multicolumn{2}{c|}{1000 kbps} 
				& \multicolumn{2}{c|}{1000 kbps} 
				& \multicolumn{2}{c|}{1000 kbps} \\ \hline
				\textbf{BW per SM} 
				& \multicolumn{2}{c|}{12.5 kbps} 
				& \multicolumn{2}{c|}{50 kbps} 
				& \multicolumn{2}{c|}{50 kbps} 
				& \multicolumn{2}{c|}{50 kbps} \\ \hline
		\end{tabular}}
		\label{table:network-packet-bandwidth}
	\end{table*}
}

\subsection{Privacy Evaluation}
\label{sec:privacy-evaluation}

In this section, we formally analyze the privacy and supporting security properties of the proposed protocol under the threat model and assumptions defined earlier. For participant anonymity, we consider probabilistic polynomial-time adversaries and follow the standard game-based methodology, in which the security property is defined through an experiment between a challenger and an adversary and is quantified by the adversary's advantage \cite{bellare2006security}. The local differential privacy guarantee is analyzed independently of computational assumptions using the mathematical definition, Gaussian mechanism, and  composition properties of differential privacy \cite{dwork2006calibrating,dwork2014algorithmic}. First, we establish the consumption-value privacy provided for each perturbed report and for the complete reporting session. Subsequently, we analyze participant anonymity and issuance--reporting unlinkability against
semi-trusted entities according to the notions of anonymity and unlinkability defined in \cite{pfitzmann2010terminology}. Finally, we discuss the supporting security properties of the protocol, including confidentiality, integrity, credential-chain validity, and token security, under the assumed security of the employed cryptographic primitives. Throughout the analysis, we distinguish privacy against semi-trusted entities from security against untrusted active and
passive adversaries.

\subsubsection{Formal Consumption-Value Privacy}
\label{sec:formal-consumption-privacy}

Temporal-based aggregation converts fine-grained consumption values into coarse-grained values, thereby reducing the level of detail disclosed to the utility provider. However, aggregation alone does not provide a formal indistinguishability guarantee. In the proposed protocol, this guarantee is provided by the local Gaussian mechanism applied independently by each smart meter before reporting its consumption value. We formalize the per-report and session-level privacy guarantees as follows.

\noindent\textbf{Lemma 1 (Per-Report LDP Guarantee).}
For a selected program $pr_s$ with $nsc_s>0$, let
$\mathcal{R}_s$ denote the local randomizer that clips the coarse-grained consumption value to $[0,C_s]$ and adds Gaussian noise with standard deviation $\hat{\sigma}_s$. If
$0<\epsilon_s<1$, $0<\delta<1$, and
$\epsilon_s=\epsilon/nsc_s$, then each perturbed report
$\tilde{c}_{t_i}\leftarrow\mathcal{R}_s(c_{t_i})$
satisfies $(\epsilon_s,\delta)$-local differential privacy.

\begin{IEEEproof}
	According to Relation
	\ref{eq:vas-coarse-sensitivity}, the $L_2$-sensitivity of the clipped coarse-grained consumption value is
	$\Delta_{2,s}=C_s$. From Relations
	\ref{eq:vas-protocol-details-baseline-noise-definition},
	\ref{eq:vas-protocol-details-applying-scale-parameter}, and
	\ref{eq:vas-effective-privacy-budget}, the applied noise standard deviation can be rewritten as
	\begin{IEEEeqnarray}{rCl}
		\hat{\sigma}_s
		&=&
		nsc_s
		\dfrac{\Delta_{2,s}\sqrt{2\ln(1.25/\delta)}}{\epsilon}
		\notag\\
		&=&
		\dfrac{\Delta_{2,s}\sqrt{2\ln(1.25/\delta)}}{\epsilon_s}.
		\label{eq:vas-privacy-evaluation-effective-noise}
	\end{IEEEeqnarray}
	This is the standard Gaussian-mechanism calibration for
	$(\epsilon_s,\delta)$-differential privacy
	\cite{dwork2006calibrating,dwork2014algorithmic}. Since the randomization is performed locally by the smart meter before the report is disclosed, the resulting mechanism satisfies
	$(\epsilon_s,\delta)$-local differential privacy for every two alternative admissible consumption values.
\end{IEEEproof}

\noindent\textbf{Theorem 1 (Session-Level Consumption-Value Privacy).}
Let $q_s=freq_s\cdot pd_s$ denote the number of reports generated during program $pr_s$. Under the conditions of Lemma 1, the complete reporting session satisfies
$(q_s\epsilon_s,q_s\delta)$-local differential privacy, provided that $q_s\delta<1$.

\begin{IEEEproof}
	By Lemma 1, every report generated during the program satisfies
	$(\epsilon_s,\delta)$-local differential privacy. Applying the basic sequential composition theorem to the $q_s$ reports yields
	\begin{IEEEeqnarray}{rCl}
		\epsilon_s^{\mathrm{session}}
		&=&
		q_s\epsilon_s
		\notag\\
		\delta_s^{\mathrm{session}}
		&=&
		q_s\delta.
		\label{eq:vas-privacy-evaluation-session-composition}
	\end{IEEEeqnarray}
Therefore, for any two possible consumption sequences containing $q_s$ reports, the probability distributions of the corresponding noisy reporting transcripts differ by at most the composed privacy bound $(q_s\epsilon_s,q_s\delta)$ \cite{dwork2014algorithmic}.
\end{IEEEproof}

The session-level bound explicitly shows the cumulative privacy loss caused by repeated reporting. Increasing $nsc_s$ decreases the effective per-report budget $\epsilon_s$ and consequently strengthens both the per-report and session-level privacy guarantees. Conversely, increasing $freq_s$ or $pd_s$ increases $q_s$ and therefore increases the accumulated privacy loss. If $nsc_s=0$, the Gaussian randomization step is omitted and no local differential privacy guarantee is provided. Moreover, any subsequent processing of $\tilde{c}_{t_i}$ does not weaken its LDP guarantee according to the post-processing property of differential privacy \cite{dwork2014algorithmic}.

\subsubsection{Participant Anonymity and Issuance--Reporting Unlinkability}
\label{sec:participant-anonymity}

Anonymity means that an adversary cannot sufficiently identify the
actor performing an action within a set of possible actors, whereas
unlinkability prevents the adversary from sufficiently relating two
items of interest \cite{pfitzmann2010terminology}. In the proposed
protocol, $ID_{pr_s}$ forms the set of possible reporting meters for
program $pr_s$. According to Relation
\ref{eq:vas-protocol-details-anonymity-setsize}, the program is executed
only when $|ID_{pr_s}|>\mathit{threshold}$. In addition to this
structural condition, we analyze participant anonymity using a
game-based indistinguishability experiment
\cite{backes2013anoa}.

\noindent\textbf{Participant-Anonymity Experiment.}
Let $\mathcal{A}$ be a probabilistic polynomial-time adversary
representing the combined view of the semi-trusted aggregator and
utility provider. The adversary selects two honest meters
$id_0,id_1\in ID_{pr_s}$ and two admissible consumption sequences
$\mathbf{c}_0,\mathbf{c}_1\in[0,C_s]^{q_s}$ for the same program.
The sequences may be selected according to any auxiliary information
available to $\mathcal{A}$. Both meters independently generate their
credential chains and complete the authenticated credential-issuance
phase. Hence, $\mathcal{A}$ observes each meter's identity, program
identifier, and blinded credential during issuance.

The challenger samples
$b\overset{\$}{\leftarrow}\{0,1\}$ and generates the complete noisy
reporting transcript of meter $id_b$ from $\mathbf{c}_b$, using an
independently generated blinding factor, credential chain, and session
pseudonym. The transcript is forwarded through the anonymous overlay
network. All report fields use fixed-length binary representations;
therefore, the message length is independent of the numerical
consumption values. After receiving its protocol view, $\mathcal{A}$
outputs a guess $b'$. Its advantage is defined as:
\begin{IEEEeqnarray}{rCl}
	\operatorname{Adv}_{\Pi,\mathcal{A}}^{\mathrm{PA}}(\lambda)
	&=&
	\left|
	\Pr[b'=b]-\dfrac{1}{2}
	\right|.
	\label{eq:vas-participant-anonymity-advantage}
\end{IEEEeqnarray}

\noindent\textbf{Theorem 2 (Value-Aware Participant Anonymity).}
Let $\bar{\epsilon}_s=q_s\epsilon_s$ and
$\bar{\delta}_s=q_s\delta$ denote the session-level LDP parameters
established in Theorem~1. Let $\mathsf{BS}$ and $\mathsf{ON}$ denote
the employed blind-signature scheme and anonymous overlay network,
respectively.

For every participant-anonymity adversary $\mathcal A$, let
$\mathcal B_1$ and $\mathcal B_2$ denote reduction algorithms that use
the success of $\mathcal A$ to attack the blindness of $\mathsf{BS}$
and the sender anonymity of $\mathsf{ON}$, respectively. Moreover,
let $\mathcal D$ denote a distinguisher that attempts to identify the
reporting meter using only the noisy consumption transcript.

For a randomly selected
$b\overset{\$}{\leftarrow}\{0,1\}$, let
\begin{IEEEeqnarray}{rCl}
	T_b
	&\leftarrow&
	\mathcal R_s^{q_s}(\mathbf c_b),
	\label{eq:vas-consumption-challenge-transcript}
\end{IEEEeqnarray}
where $\mathcal R_s^{q_s}$ denotes the session-level local
randomization mechanism. The consumption-value distinguishing
advantage of $\mathcal D$ is defined as
\begin{IEEEeqnarray}{rCl}
	\operatorname{Adv}_{\mathcal R_s,\mathcal D}^{\mathrm{CV}}
	(\bar{\epsilon}_s,\bar{\delta}_s)
	&=&
	\left|
	\Pr[\mathcal D(T_b)=b]-\dfrac{1}{2}
	\right|.
	\label{eq:vas-consumption-distinguishing-advantage}
\end{IEEEeqnarray}
According to Theorem~1, this advantage is limited by the
session-level
$(\bar{\epsilon}_s,\bar{\delta}_s)$-LDP guarantee.

Assume that $\mathsf{BS}$ satisfies blindness
\cite{chaum1983blind}, $\mathsf{ON}$ provides sender anonymity against
the considered adversary \cite{ren2016anonymous}, and all
session-specific random values are generated independently. Then,
\begin{IEEEeqnarray}{rCl}
	\operatorname{Adv}_{\Pi,\mathcal A}^{\mathrm{PA}}(\lambda)
	&\leq&
	\operatorname{Adv}_{\mathsf{BS},\mathcal B_1}^{\mathrm{blind}}
	(\lambda)
	+
	\operatorname{Adv}_{\mathsf{ON},\mathcal B_2}^{\mathrm{SA}}
	(\lambda)
	\notag\\
	&&+
	\operatorname{Adv}_{\mathcal R_s,\mathcal D}^{\mathrm{CV}}
	(\bar{\epsilon}_s,\bar{\delta}_s)
	+
	\operatorname{negl}(\lambda).
	\label{eq:vas-participant-anonymity-bound}
\end{IEEEeqnarray}
The first three terms represent identity inference through credential
issuance, the communication path, and the noisy consumption values,
respectively. The negligible term accounts for accidental collisions among
independently generated session-specific values, including pseudonyms,
credential seeds, and blinding factors.
\begin{IEEEproof}
	The proof follows a hybrid argument that separates the possible
	sources of identity information. If $\mathcal A$ identifies the
	reporting meter by linking the presented credential-signature pair to
	an authenticated issuance session, $\mathcal B_1$ can use this result
	to violate the blindness of $\mathsf{BS}$. If $\mathcal A$ identifies
	the reporting meter from the communication path, $\mathcal B_2$ can
	use this result to violate the sender anonymity of $\mathsf{ON}$.
	
	After excluding credential linkage and network-source information,
	any remaining distinction between $id_0$ and $id_1$ must be obtained
	from their noisy consumption transcripts. This distinction is
	represented by
	$\operatorname{Adv}_{\mathcal R_s,\mathcal D}^{\mathrm{CV}}$ and is
	limited by the session-level LDP guarantee established in Theorem~1.
	All remaining differences caused by accidental equality among fresh
	session-specific values occur with negligible probability. Combining
	these sources establishes
	Relation~\ref{eq:vas-participant-anonymity-bound}.
\end{IEEEproof}

This result considers the case where the two challenge meters have different consumption patterns and the adversary has auxiliary
information about them. The blind-signature scheme and anonymous overlay network limit identity leakage during credential issuance and
report transmission, while the LDP mechanism limits inference from the reported consumption values. Reports generated within the same session
can be linked through $uuid_{\omega}$ and the credential hash chain for verification purposes; however, they cannot be linked to the meter's
real identity.

\subsubsection{Protocol Security under Standard Cryptographic Assumptions}
\label{sec:protocol-security}

Assuming the security of the employed public-key encryption, digital signature, blind-signature, MAC, and hash schemes, the proposed protocol provides the supporting security properties required under the defined threat model. Encryption protects the confidentiality of the protocol messages against untrusted adversaries, while digital signatures and MACs enable the legitimate entities to detect unauthorized message modifications. The utility provider's signature
on the last credential validates the initial anonymous credential, and the preimage resistance of $H$ prevents an adversary from generating the subsequent credentials without knowing the valid credential chain. Furthermore, each token is uniquely identified and signed by the utility provider, while its redemption status is maintained in the database to prevent forgery and double spending. These properties hold under the trusted-meter and semi-trusted-entity assumptions defined in the threat model.

\section{Conclusion}
In this paper, we introduced a lightweight incentive-based privacy-preserving smart metering protocol for value-added services.
The proposed protocol enables smart meters to report coarse-grained or noisy coarse-grained consumption values anonymously while receiving
incentives for their participation. Hash-chain credentials together with blind digital signatures enable the utility provider to authenticate
successive anonymous reports while separating authenticated program participation from subsequent reporting, thereby supporting  issuance--reporting unlinkability. Temporal-based aggregation reduces the granularity of disclosed consumption values, while the local differential
privacy mechanism provides a formal privacy guarantee when noise is enabled. Specifically, each perturbed report satisfies
$(\epsilon_s,\delta)$-local differential privacy, and a complete program with $q_s=freq_s\cdot pd_s$ reports satisfies  $(q_s\epsilon_s,q_s\delta)$-local differential privacy under the stated conditions. Pseudonyms and the anonymous overlay network further protect
participant identity, and our formal analysis establishes participant anonymity and issuance--reporting unlinkability against the considered
semi-trusted entities. Encryption, digital signatures, and message authentication codes are additionally employed to provide the required
confidentiality and integrity during protocol execution.

The protocol allows customers to select programs with different reporting frequencies, durations, noise levels, and rewards, enabling the utility
provider to construct privacy-preserving datasets with adjustable data utility for value-added services. The performance results also demonstrate
the feasibility of the proposed approach. With a 2048-bit RSA key size, a 7-day program duration, and four reports per day, the complete execution of the protocol takes approximately $3.36$ seconds, including about $3.03$ seconds at the smart meter and $0.33$ seconds at the utility provider.
For the resulting 28 reports, this corresponds to an amortized overall computational overhead of approximately $0.12$ seconds per report. Under
the same setting, the complete protocol consumes approximately $4.5$ MB of memory on average. These results, together with the established
consumption-value privacy, participant anonymity, and issuance--reporting unlinkability properties, show that the proposed protocol can provide privacy-preserving data collection for value-added services with practical computational and memory requirements.


%





\ifCLASSOPTIONcaptionsoff
  \newpage
\fi





\bibliographystyle{IEEEtran}
\bibliography{IEEEabrv,Bibliography}

\vfill


\end{document}